\documentclass{article}

\PassOptionsToPackage{numbers, sort&compress}{natbib}
\usepackage[final]{neurips_2023}
\usepackage{longtable}
\usepackage{hyperref}

\usepackage[utf8]{inputenc} %
\usepackage[T1]{fontenc}    %
\usepackage{hyperref}       %
\usepackage{url}            %
\usepackage{booktabs}       %
\usepackage{amsfonts}       %
\usepackage{nicefrac}       %
\usepackage{microtype}      %
\usepackage{xcolor}         %
\usepackage[pdftex]{graphicx}
\usepackage[textsize=tiny]{todonotes}
\usepackage{needspace}

\usepackage[scaled=0.86]{helvet} %

\usepackage{chemformula}
\usepackage{xspace}
\usepackage{bm}
\usepackage{enumitem}
\usepackage{wrapfig}
\usepackage{placeins}
\usepackage{float}  %
\usepackage{makecell}
\usepackage{siunitx}
\usepackage[ruled,noend,vlined,algosection]{algorithm2e}

\SetCommentSty{mycommfont}
\SetKwComment{tcp}{  $\triangleright$ }{}
\newcommand{\nlother}[1]{\refstepcounter{AlgoLine}\nlset{\theAlgoLine\rlap{#1}}}
\SetKwFor{ParFor}{for}{do (in parallel)}{end ParFor}

\newcommand{\ourModel}{\texttt{SCARF}\xspace}
\newcommand{\ourModelOne}{\texttt{SCARF-Thread}\xspace}
\newcommand{\ourModelTwo}{\texttt{SCARF-Weave}\xspace}

\newcommand{\ourModelOneShort}{\texttt{Thread}\xspace}
\newcommand{\ourModelTwoShort}{\texttt{Weave}\xspace}

\newcommand{\cfmModel}{\texttt{CFM-ID}\xspace}

\newcommand{\nistData}{\texttt{NIST20}\xspace}
\newcommand{\gnpsData}{\texttt{NPLIB1}\xspace}
\newcommand{\MAGMA}{\texttt{MAGMa}\xspace}

\newcommand{\mol}{\mathcal{M}}
\newcommand{\mlp}{\textsf{MLP}}
\newcommand{\bsigma}{\bm{\sigma}}

\newcommand{\reals}{\mathbb{R}}
\newcommand{\chemSpace}{\mathcal{X}}
\newcommand{\preForm}{\bm{\mathcal{F}}}
\newcommand{\preFormIdxd}[1]{\mathcal{F}_{#1}}
\newcommand{\prodForm}[1]{\bm{f}^{#1}}
\newcommand{\prodFormIdxd}[2]{f^{#1}_{#2}}
\newcommand{\inten}[1]{y^{#1}}
\newcommand{\numProds}{n}
\newcommand{\numElements}{e}
\newcommand{\setOfProdForm}{\{ \prodForm{i} \}_{i=1}^{\numProds}}

\newcommand{\setOfProdsAndIntens}{\left\{ (\prodForm{i}, \inten{i}) \right\}_{i=1}^{\numProds}}
\newcommand{\predBinnedSpec}{\hat{\bm{s}}}
 \newcommand{\binnedSpec}{\bm{s}}

 \newcommand{\codeUrl}{at \url{https://github.com/samgoldman97/ms-pred}}

\definecolor{modelOneBlue}{HTML}{F8971D} %
\definecolor{modelTwoOrange}{HTML}{479394}
\definecolor{jbcommentColor}{HTML}{ff66ff}
\definecolor{sgCommentColor}{HTML}{00cc66}

\usepackage{xr}

\title{Prefix-Tree Decoding for Predicting Mass Spectra from Molecules}

\author{%
  Samuel Goldman \\
    \parbox{18em}{\centering Computational and Systems Biology}\\
  MIT\\
  Cambridge, MA 02139 \\
  \texttt{samlg@mit.edu} \\
  \And
  John Bradshaw \\
  \parbox{18em}{\centering Chemical Engineering} \\
  MIT \\
  Cambridge, MA 02139 \\
  \texttt{jbrad@mit.edu} \\
  \AND
  Jiayi Xin \\
  \parbox{18em}{\centering 
  Statistics and Actuarial Science}\\
The University of Hong Kong\\
Pokfulam, Hong Kong\\
  \texttt{xinjiayi@connect.hku.hk} \\
  \And
  Connor W.~Coley \\
  Chemical Engineering\\
   \parbox{18em}{\centering Electrical Engineering and Computer Science} \\
  MIT\\
  Cambridge, MA 02139\\
  \texttt{ccoley@mit.edu} \\
}

\begin{document}

\maketitle

\begin{abstract}
Computational predictions of mass spectra from molecules have enabled the discovery of clinically relevant metabolites. However, such predictive tools are still limited as they occupy one of two extremes, either operating  (a) by fragmenting molecules combinatorially with overly rigid constraints on potential rearrangements and poor time complexity or (b) by decoding lossy and nonphysical discretized spectra vectors.
In this work, we use a new intermediate strategy for predicting mass spectra from molecules by treating mass spectra as sets of molecular formulae, which are themselves multisets of atoms. After first encoding an input molecular graph, we decode a set of molecular subformulae, each of which specify a predicted peak in the mass spectrum, the intensities of which are predicted by a second model. Our key insight is to overcome the combinatorial possibilities for molecular subformulae by decoding the formula set using a prefix tree structure, atom-type by atom-type, representing a general method for ordered multiset decoding. We show promising empirical results on mass spectra prediction tasks. 
\end{abstract}

\section{Introduction}%
\label{sec:intro}%
\begin{wrapfigure}[12]{r}{7.5cm}
  \centering
  \vspace{-2.2em}\includegraphics[width=7.5cm]{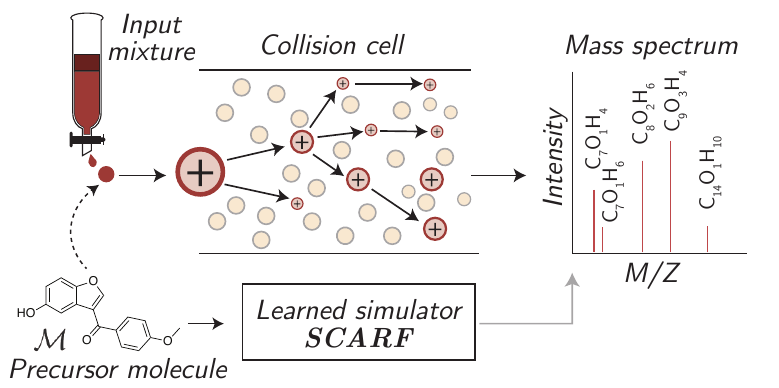}
  \label{fig:scarfTeaser}
    \vspace{-1.5em}
  \caption{
    Tandem mass spectrometers measure fragmentation patterns of  molecules, resulting in characteristic peaks that are indicative of their structure. \ourModel simulates these fragmentation patterns \textit{in silico}. 
  }
\end{wrapfigure}
As the primary tool to discover unknown small molecule 
structures from biological samples, tandem mass spectrometry (MS/MS) experiments have enabled the identification of numerous important molecules implicated in health and disease \cite{wishart_metabolomics_2019, quinn_global_2020, bauermeister2022mass}. Tandem mass spectrometers are capable of isolating, fragmenting, and measuring the resulting fragment masses of small molecules from a sample, producing a signature (a mass spectrum) for each detected molecule (Figure~\ref{fig:scarfTeaser}, top).

Computationally predicting mass spectra from molecules \emph{in silico} (Figure~\ref{fig:scarfTeaser}, bottom) is thus a longstanding and important
challenge. Not only does this assist practitioners in better understanding the
fragmentation process, but it also enables the
identification of molecules from newly observed spectra by comparing an observed
spectrum to virtual spectra generated from a
database of candidate molecules. While a large library of empirical mass spectra could theoretically serve the same purpose, the size of such libraries is limited by the slow and expensive process of acquiring pure chemical standards and measuring their spectra, motivating computational prediction.

We argue that there are three core, interrelated desiderata for a forward molecule-to-spectrum simulation model, or ``spectrum predictor''.
An ideal spectrum predictor should be 
(i) \emph{accurate}, being able to predict the exact set of fragment masses and intensities with a precision comparable to experimental measurements;
(ii) \emph{physically inspired}, to avoid making physically nonsensical (``invalid'') suggestions and to provide interpretations of the chemical species responsible for each peak for the benefit of human expert chemists;
and (iii) \emph{fast}, such that it is computationally inexpensive to predict spectra for many (e.g., millions) hypothetical molecules. %

\begin{figure}[t]
  \centering
  \includegraphics[width=\textwidth]{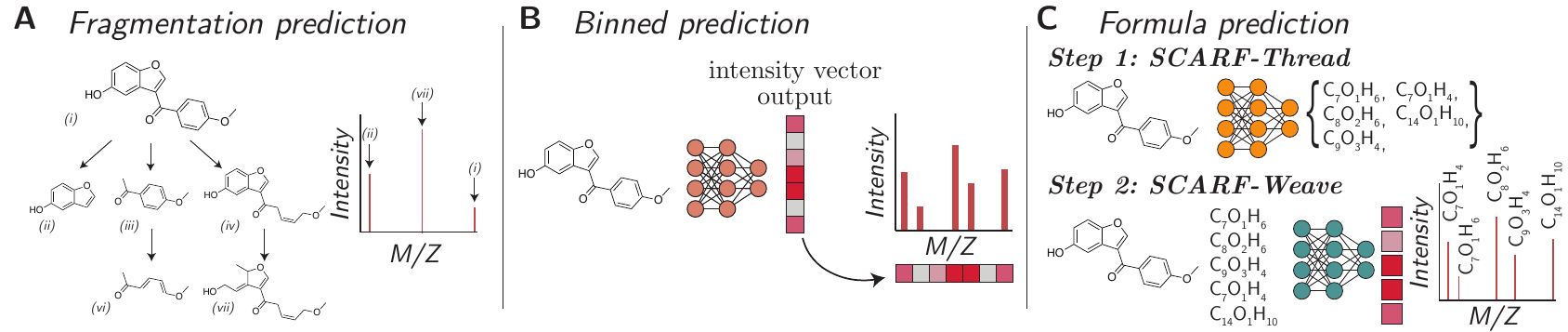}
  \label{fig:background}
  \caption{Overview of various approaches to spectrum prediction. \textbf{A.} Fragmentation prediction approaches use
  heuristics and scoring rules to break down the molecule into fragments and
  their associated intensities.
  \textbf{B.} 
  Binned prediction approaches discretize the possible mass-to-charge values and
  predict intensities for each possible bin. \textbf{C.} Formula prediction approaches predict spectra as sets of molecular formulae and intensities.
  Our model \ourModel utilizes a two stage approach, first by predicting the product
  formulae present (constrained by the precursor formula), which defines the
  x-axis locations of the peaks, before secondly assigning intensities to these
  formulae (defining the peaks' y-axis values).}
\end{figure}

Unfortunately, many existing spectrum predictors do not meet these criteria.
Methods to date have tended to follow one of two approaches: (a) physically motivated fragmentation
approaches or (b) molecule-to-vector (or ``binned'') approaches (Figure
\ref{fig:background}A-B). Fragmentation approaches (e.g.,
\citep{allen_competitive_2015,wolf_silico_2010, ridder_automatic_2014,
Gasteiger1992-aj}; Figure \ref{fig:background}A) take an input molecule and suggest bonds that
may break, creating fragments that are scored by ML algorithms or curated
rulesets.  While interpretable, these methods are often slow and
restrictive; certain mass spectrum peaks are generated by complex chemical rearrangements
within the collision cell that cannot be approximated by bond breaking alone. That is,  the bonds in observed fragments are not a subset of those in the original
molecule \cite{demarque2016fragmentation, chen2001rearrangement}.  On the other
hand, binned prediction approaches (e.g., \citep{wei_rapid_2019,
zhu_using_2020, young_massformer_2021}; Figure \ref{fig:background}B) are less physically grounded, using neural networks to directly learn a mapping from molecules to
vectors representing discretized versions of the spectra.
These methods, while fast, lack interpretability and due to discretization have a mass precision lower than that of most modern spectrometers, limiting their accuracy.

We propose to address the shortcomings of previous work by predicting mass spectra from molecules at the level of molecular formulae (e.g., C$_x$N$_y$O$_z$H$_w$...) and introduce a new method, Subformulae Classification for Autoregressively Reconstructing Fragmentations (\ourModel) to do so. Because the molecular formula for each input molecule is known, each subformula in the predicted set of peaks is constrained to contain a subset of the atoms in the original formula.
Our primary contributions are:
\begin{itemize}%
    \item posing mass spectrum prediction as a two step process: first generating the set of molecular formulae for the fragments, then associating these formulae with intensities;
    \item overcoming the combinatorial subformula option space by learning to generate formula prefix trees;
    \item demonstrating the empirical benefit of \ourModel in predicting experimental mass spectra quickly and accurately using two separate datasets, providing a benchmark for future work.
\end{itemize}

\FloatBarrier

\section{Background}
\label{sec:background}

We provide a short introduction of tandem mass spectrometry suitable for a
general machine learning audience, detail  previous approaches to modeling this process
as they relate to our proposed approach \ourModel, and explain how such tools can be
utilized to discover molecules from new spectra. We refer interested readers to
\cite{hufsky2014computational} for further details on the physical process of mass spectrometry.

\subsection{Tandem mass spectrometry}

Tandem mass spectrometers (MS/MS) measure fragmentation patterns of molecules in
a multi-stage process.  The input to the process is a solution containing a \emph{precursor} molecule,
$\mol \in \chemSpace$, associated with a molecular formula, $\preForm$, defining
the counts of each element present; for instance $\preForm =  \ch{C16O4H12}$ for
the precursor molecule shown in Figure~\ref{fig:scarfTeaser}.  The precursor
molecule is first ionized (i.e., made charged), often by bonding or associating with an \emph{adduct} (e.g., a proton, \ch{H+}) present in the solution. 
The charged product is then measured by a mass analyzer (MS1), where its mass-to-charge
ratio ($m/z$) is measured.

This precursor ion is then filtered into a \emph{collision cell}. Here, through
interactions with an inert gas, the precursor ion is broken down into a set of
one or more \emph{product ions}, each of which is associated with a new chemical
formula; for example, one might be $\prodForm{1} = \ch{C7OH4}$ for the process
shown in Figure~\ref{fig:scarfTeaser}.  Finally, this set of product ions is
measured by a second mass analyzer (MS2), along with the set of their intensities, $\inten{i} \in \reals^+$ (i.e.,
their relative frequencies over several repetitions of this process), creating for each ion what is
referred to as a \emph{peak}.  The collection of all peaks makes up a molecule's
\emph{mass spectrum}, and is commonly represented as a plot of intensities versus
$m/z$ (Figure \ref{fig:scarfTeaser}, right).

\subsection{Predicting mass spectra from molecules (spectrum predictors)}

\paragraph{Fragmentation prediction.}
A complex but physically grounded strategy is to model the bond
breakage processes occurring in the collision cell
(Figure \ref{fig:background}A).  Examples include MetFrag
\cite{wolf_silico_2010}, MAGMa \cite{ridder_automatic_2014}, and CFM-ID
\cite{allen_competitive_2015}, which recursively fragment molecules (either
bond or atom removals) to generate fragment predictions. These methods combine
expert rules and local scoring methods to enumerate molecular fragmentation
trees to predict spectra. CFM-ID \cite{allen_competitive_2015} learns subsequent
fragmentation transition probabilities between fragments with an expectation
maximization algorithm to determine intensities at each fragment. Rule-based
methods and full tree enumeration reduce the flexibility of these approaches,
and along with the inherent ambiguity in the fragmentation process, limit this strategy's overall accuracy and speed. 

\paragraph{Binned prediction.} 
An increasingly popular and straightforward approach to spectra prediction is to map molecules
to discretized 1D mass spectra from either molecular fingerprint
\cite{wei_rapid_2019} or graph inputs \citep{zhu_using_2020,
young_massformer_2021} (Figure \ref{fig:background}B). Specifically, these
methods divide the $m/z$ axis into fixed-width ``bins'' and predict an aggregate
statistic of the peaks found in each bin (such as their maximum or summed intensity).
While more flexible and end-to-end than fragmentation-based approaches, these methods do not impose the same
physical constraints or shared information across fragments, making them less interpretable and susceptible to making invalid predictions. Further, discretizing the input spectrum inherently restricts the precision of such models compared to exact-mass predictions.

\paragraph{Formula prediction.} 

We introduce the strategy of predicting spectra at the level of molecular
formulae, an intermediate between binned and fragmentation prediction (Figure \ref{fig:background}C).
Simultaneous to our work, two groups have separately explored formula prediction strategies \cite{murphy2023efficiently, zhu2023rapid}.
However, to generate plausible subformulae candidates, they either generate a
fixed vocabulary of formulae 
\citep{murphy2023efficiently} or restrict their model to 
molecules under 48 atoms for exhaustive enumeration \citep{zhu2023rapid}, which is smaller than many compounds of interest. We overcome the combinatorial problem of formula generation using
prefix trees, allowing our method to scale and eliminating the need for large, fixed vocabularies. 

\subsection{Mass spectrum libraries}

One important use of spectrum predictors is in building large \emph{in silico} libraries of molecule spectra to augment the small size of existing, experimentally derived databases (on the order of $10^4$) which are expensive to curate. 
These spectra libraries are then leveraged downstream in different ways, for example for training molecular property predictors directly from mass spectra \cite{voronov2022ms2prop}.
Another common application of spectra libraries is to infer an unknown molecule's structure from a newly observed spectra -- a particularly hard problem, with only 13\% of spectra measured from clinical samples identifiable using current elucidation tools \citep{bittremieux2022critical}.
In this problem, spectra libraries are used as part of a process called \emph{retrieval}: The newly observed spectra is compared with the existing spectra in the library using a fixed or learned spectral distance function, such as cosine distance  \citep{huber_matchms-processing_2020, huber_ms2deepscore_2021, Bittremieux2022-nk}, and the molecules associated with the closest spectra are returned as possible matches. 
In practice, the retrieval
process is constrained to choosing among \emph{isomers}
(i.e., molecules with the same molecular formula, and therefore molecular weight, but with different bond
configurations) due to the high resolution of modern mass
spectrometers (i.e., absolute errors on the order of $10^{-4}$ to $10^{-3}$ $m/z$ for MS1 measurements) \citep{duhrkop_sirius_2019,
  ludwig_database-independent_2020, xing2023buddy}.

Given the varied use cases of spectra libraries, we focus on evaluating spectrum predictors in terms of both (a) their prediction accuracies (\S\ref{sec:spectra-prediction}), using metrics such as ``cosine similarity'', and (b)  their use in generating virtual spectral libraries to assist with retrieval (\S\ref{sec:retrieval}).

\section{Model}
\label{sec:methods}

Here, we describe our model, \ourModel, for predicting mass
spectra from precursor molecules via first predicting subformulae of the precursor molecule, referred to as \emph{product formulae}.  Building upon the notation introduced in the previous
section, we continue to denote precursor molecules\footnote{We model
  and discuss uncharged molecules and formulae, despite mass
  spectrometry measuring the masses of adduct \emph{ions}. In
  practice, we reduce all molecules to uncharged candidates by simply
  shifting all the spectra weights by the $m/z$ of their respective
  adducts, which we assume to be equal to the (known) adduct of the parent molecule.} as $\mol \in \chemSpace$, and their associated formula
vector as $\preForm \in \mathbb{N}_0^\numElements$, defining at each
position, $j \in \{1, \ldots, e\}$, the count of each possible
chemical element present, $\preFormIdxd{j}$ (with zero indicating none
of that chemical element is present).  Likewise, we define the set of
$n$ product formulae as $\setOfProdForm$, and associate with each an
intensity, $\inten{i}$.  Note that
the mass\footnote{We assume singly charged adducts (as is common practice, \citep{allen_competitive_2015}),
  such that masses and mass-to-charge ratios are interchangeable.}  corresponding
to a given formula (and, as such, the x-axis
location of the peak on a mass spectrum) is determined
deterministically from the counts of each elements present.

At a high level, \ourModel generates mass spectra through the composition of two learned functions:
\begin{equation}
\begin{aligned}
 \setOfProdsAndIntens = & \;
                             \textcolor{modelTwoOrange}{g_{\theta}^{\ourModelTwoShort}} \Big(
                             \textcolor{modelOneBlue}{ g_{\theta}^{\ourModelOneShort}}\left( \mol \right), \mol \Big),
\end{aligned}
\label{eqn:scarf}
\end{equation}%
first mapping from the original molecule to a set of product formulae,
$  \textcolor{modelOneBlue}{g_{\theta}^{\ourModelOneShort}}: \mol \mapsto \setOfProdForm $,
and then mapping from this set of formulae (and the original molecule) to the respective intensities,
$ \textcolor{modelTwoOrange}{g_{\theta}^{\ourModelTwoShort}}: (\setOfProdForm, \mol) \mapsto \setOfProdsAndIntens $
. The particularities of both functions are described in detail below. The specific architectures and hyperparameters used are deferred to the appendix;
model code can be found \codeUrl.

\subsection{\textcolor{modelOneBlue}{\ourModelOne} : Generating product formulae via generating prefix trees}

\begin{figure*}[t]
  \centering
  \includegraphics[width=\textwidth]{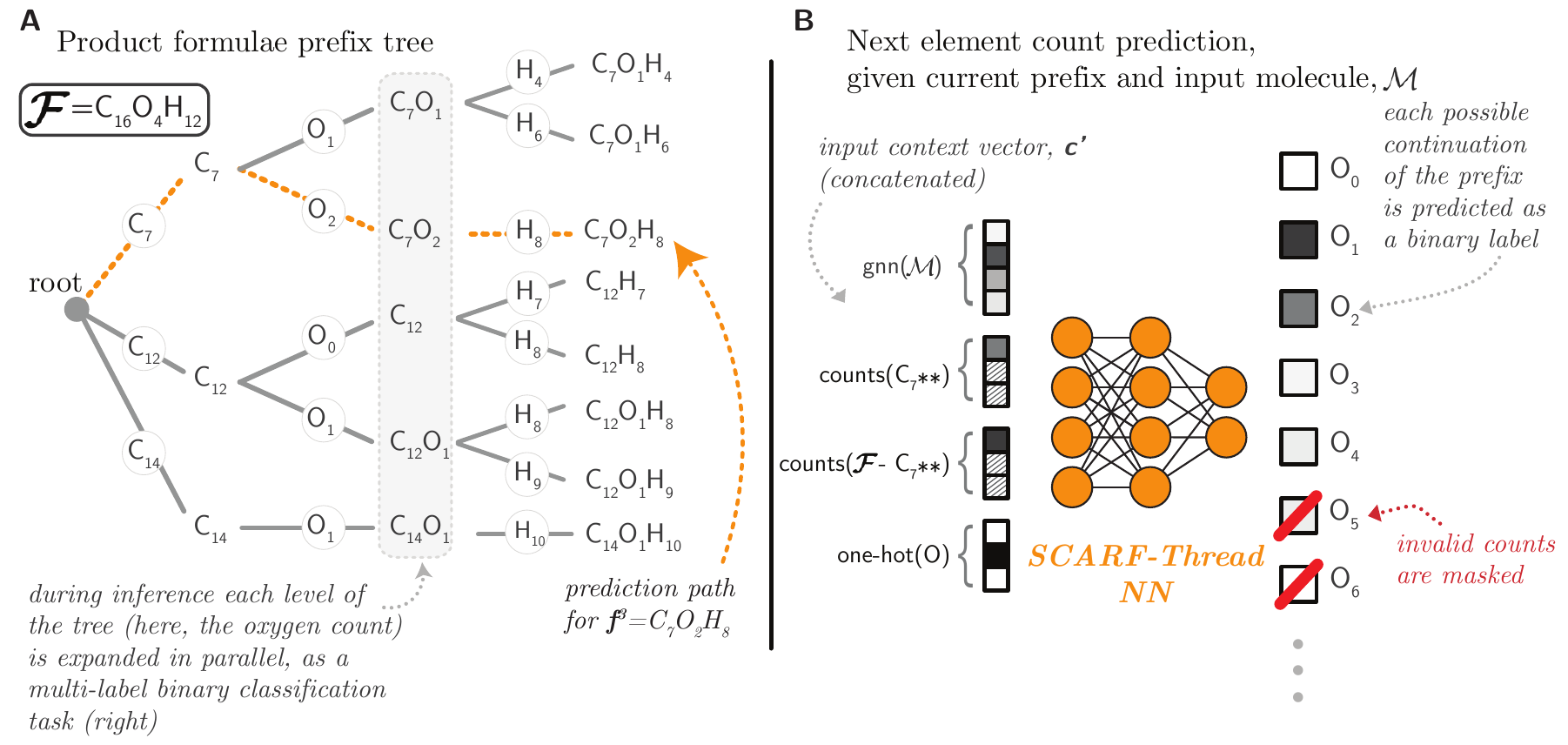}
  \caption{
    Illustration of the \ourModelOne architecture.
    \textbf{A.} The formulae of the product fragments can be represented using a
    prefix tree.  \ourModelOne predicts this tree for new
    molecules at test time.  It does so by expanding each node at a given depth
    in parallel, treating the counts of subsequent elements as dependent only on
    the counts of elements predicted so far (i.e., the prefix) and the original molecular structure.
    \textbf{B.} The \ourModelOne predictive task at the \ch{C7} node from the
    prefix tree diagram shown in A. Here the network takes as input (i) an embedding of
    the overall molecule; (ii) a vector representing the counts of each element in
    the prefix so far (counts yet to be predicted are represented using a
    special token), (iii) the difference of the counts predicted so far from the
    precursor molecule, and (iv) a one-hot representation of the element for which
    the counts are currently being predicted.  The network predicts which counts
    are valid next nodes in the prefix tree (where counts that are greater than
    those in the original precursor molecular formula are automatically masked out as invalid).
    See also Alg.~\ref{alg:modelOne}.
   }
   \label{fig:scarfModelOne}
\end{figure*}

\ourModelOne is tasked to learn a mapping to the set of product formulae, $\setOfProdForm$, given the original molecule.
Naively, one might try to define this model autoregressively,
predicting the set formula by formula, chemical element by chemical element.
However, such an approach soon runs into a number of problems as (i) the
predictions are not invariant to set and ordering permutations; (ii) the time complexity of prediction would scale poorly, being proportional to both the number of elements and number of product formulae (i.e., $\mathcal{O}(\numElements \times \numProds )$); %
and (iii) the predictions would likely contain duplicates. %

We therefore take a different approach using the insight that the set of all product formulae can be compactly represented as a prefix tree (Figure~\ref{fig:scarfModelOne}A).  In this tree, edges at a given depth represent valid counts of a particular chemical element, which are often identical across multiple product formulae (shown in the circles). By following each path from the root node to the different leaf nodes, we can reconstruct
each product formula (as the orange dashed path does for a single product formula).

We thus propose \ourModelOne as an autoregressive generator to define a probability distribution over such a prefix tree (Alg.~\ref{alg:modelOne}).  We assume that each
product formula is a subset of the precursor formula, meaning that the precursor formula sets
an upper bound on the maximum number of each element\footnote{While it is possible for fragments to
  fuse together, potentially taking the count of a chemical element over the number in the original 
  precursor formula, we postpone the extension to modeling such rare events to future work.}.  At each node in the tree
(corresponding to a prefix $\prodForm{'}_{<j}$), we pose the prediction of the set of child nodes
(corresponding to the set of valid counts of the subsequent element) as a multi-label binary classification
problem (Figure~\ref{fig:scarfModelOne}B). Concretely, we use a neural network
module for this task, giving it as input a context vector representing the node being expanded:
\begin{equation}
  \bm{c}' = [
     \textsf{gnn}(\mol), 
     \textsf{counts}(\prodForm{'}_{<j}), 
     \textsf{counts}(\preForm - \prodForm{'}_{<j}),  
     \textsf{one-hot}(j)],
   \label{eqn:contxt}
\end{equation}
where $\textsf{gnn}(\cdot)$ specifies a neural encoding of the molecular graph
(\S\ref{sec:mol_enc}), $\textsf{counts}(\cdot)$ specifies a count-based encoding of the associated
prefix (\S\ref{sec:form_reps}), and $\textsf{one-hot}(\cdot)$ specifies a one-hot encoding of the node's depth (or
equivalently, which element the predicted count is for). 
In our experiments, we use a fixed ordering of the chemical elements (\S \ref{sec:si_datasets}), but optimizing or even learning the tree construction order could be carried out \citep{Vinyals2016-zz}.

\paragraph{Formulae as differences.}
Following \citet{wei_rapid_2019}, we find it helpful to not only parameterize product formulae
in terms of their element counts, but also in terms of the elements that they have lost, i.e., their
\emph{difference} from the precursor formula. On the input side, this is already covered by
including in the context vector a count-based embedding of the prefix formula minus the product formula
($\textsf{counts}(\preForm - \prodForm{'}_{<j})$). However, on the output side this is achieved by
combining the probabilities of a ``forward'' and a ``difference'' network:
\begin{equation}
  p(\prodFormIdxd{'}{j} = a | \prodForm{'}_{< j}, \mol) =  \bm\alpha_a  \bsigma\left( \mlp^F(\bm{c}') \right)_a + (\bm1-\bm\alpha)_a \bsigma \left(  \mlp^D(\bm{c}') \right)_{\preFormIdxd{j} - a} ,
 \label{eqn:fordiff}
\end{equation}
where $\mlp^F(\cdot)$ and  $\mlp^D(\cdot)$ specify multi-layer perceptrons (MLPs) for predicting the probability of
observing a count of $a$ and a loss of $\preFormIdxd{j} - a$ atoms respectively; $\bm\alpha$ is a variable (output from a third, unshown network) deciding how to weight these predictions; and $\bsigma(\cdot)$ is the element-wise sigmoid function.

\subsection{\textcolor{modelTwoOrange}{\ourModelTwo}: Predicting intensities given product formulae}

\begin{wrapfigure}[20]{l}{7cm}
  \centering
    \vspace{-1em}
  \includegraphics[width=7cm]{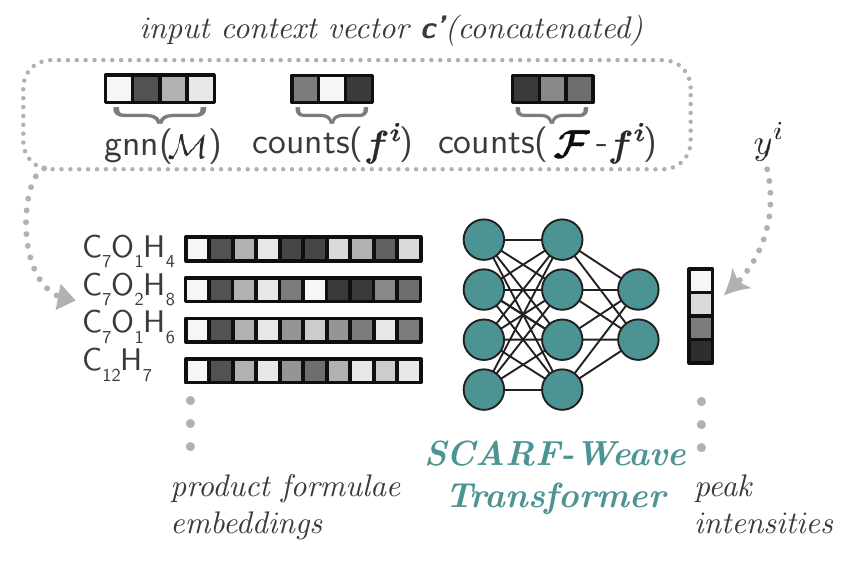}
  \caption{
    The \ourModelTwo network, which takes in the product formulae (e.g., predicted by \ourModelOne) and
    predicts their intensities.
    We use a Set Transformer architecture \citep{lee2019set}, such that our model takes in the details of the other product formulae present when predicting intensities.
  } 
  \label{fig:scarfModelTwo}

\end{wrapfigure}

Given the product formulae outputs from \ourModelOne, \ourModelTwo predicts corresponding intensities at each formula.  This is a set-to-set problem, well suited for any
equivariant set2set architecture \citep[\S3.1]{Zaheer2017-jj}.  In our experiments,
we use a Set Transformer \cite{lee2019set,Vaswani2017-se}, which enables the model to consider all the formulae present in the mass spectrum (and their possible interactions) when
predicting final intensities.

We choose to represent formula in the set similarly to the context vectors used in \ourModelOne.
For each input, we concatenate a vector embedding of the initial molecular graph with count-based embeddings
of the product formula and its difference from the precursor formula (Figure~\ref{fig:scarfModelTwo}).
We again defer the particularities of the embedding functions to the Appendix (\S \ref{sec:model_details}).\\

\subsection{Training and inference}
\label{sec:train}

Provided with a dataset of molecules and formula-labeled mass spectra, we could train the two components of
\ourModel separately. However, in practice we find it beneficial to first train \ourModelOne and then
train \ourModelTwo on its outputs so that the distribution the latter model sees is the same at training and prediction time. \ourModelTwo is trained using a cosine loss (\S \ref{sec:modelTwo}), as this most closely resembles the ``retrieval'' setting (\S~\ref{sec:retrieval}).

\ourModelOne is trained using the binary cross entropy losses associated with the multi-label
classification tasks at each non-leaf node in the prefix tree. We use teacher forcing, i.e., we train on each level of the tree in parallel by conditioning on the ground-truth set of prefixes at each stage.
In our experiments, when generating the set of product formulae from this model we always pick the top 300.
Empirically, we find that this provides better performance than picking 
a variable number based on a likelihood threshold. %

\section{Experiments}
\label{sec:experiments}

We evaluate \ourModel on spectra prediction (\S~\ref{sec:spectra-prediction}) and molecule identification in a retrieval task (\S~\ref{sec:retrieval}).

\subsection{Dataset}
\label{sec:datasets}

 We train and validate \ourModel on two libraries: a gold standard commercial tandem mass spectrometry dataset, \nistData \citep{noauthor_tandem_nodate}, as well as a more heterogeneous public dataset, \gnpsData, extracted from the GNPS database \cite{wang_sharing_2016} by \citet{duhrkop_systematic_2021} and subsequently processed by \citet{Goldman2023annotating}. We prepare both datasets by extracting and preprocessing spectra, as well as filtering to compounds that (a) are under 1,500 Da (i.e., typically under 100 heavy atoms), (b) only contain  predefined elements, and (c) are only charged with common positive-mode adduct types (\S\ref{sec:datasets}).

Overall, \nistData contains 35,129 total spectra with 24,403 unique structures, and 12,975 unique molecular formulae; \gnpsData contains 10,709 spectra, 8,553 unique structures and 5,433 unique molecular formulae. Both datasets are evaluated using a structure-disjoint 90\%/10\% train/test split with 10\% of training data held out for validation, such that all compounds in the test set are not seen in the train and validation sets.

\paragraph{Annotating spectra.}

We emphasize that \ourModel can be trained with any
product formula annotations, which can be labeled \cite{noauthor_tandem_nodate} or inferred with varied computational strategies \cite{duhrkop_sirius_2019}.
Herein, we utilize the \MAGMA algorithm \cite{ridder_automatic_2014}. In brief, for a given
molecule-spectrum pair in the training dataset, the molecule is combinatorially fragmented at each atom up to a depth of 3 breakages to create sub-fragments. This creates a bank of possible molecular formulae, and each peak in the spectrum is assigned to its nearest possible formula within a mass difference of 20
parts-per-million.

\subsection{Spectra prediction}
\label{sec:spectra-prediction}

\paragraph{Predicting product formulae (\ourModelOne).}

\ourModelOne is trained and used to reconstruct prefix trees and evaluated by
its ability to recover the ground truth product formula set. The set of generated
product formulae is rank-ordered by the probability of each product formula and
filtered to the top $k$ predicted product formulae. The fraction of ground truth
formulae (22.29 peaks on average in \nistData) contained in the top k set is computed as \emph{coverage}.

We compare coverage achieved by \ourModelOne to several baselines: (i)
\cfmModel~\cite{allen_competitive_2015}, a fragmentation based approach (\S~\ref{sec:cfm}); (ii) a
random baseline that samples product formulae from a uniform distribution; (iii) a frequency baseline, which
ranks product formulae by the frequency the product formula candidate (or
product formula difference) appears in the training set;  (iv) an LSTM autoregressive neural network baseline (\S~\ref{sec:autoregr}) that is trained to predict molecular formula vectors in sequence from highest to lowest intensity; and two model
ablations, (v) \ourModelOne-D  and (vi) \ourModelOne-F, which only make
uni-directional elements difference or forward predictions of element counts respectively (i.e., $\bm{\alpha}$ in Equation~\ref{eqn:fordiff} is fixed to $\bm0$ for (v) and $\bm{1}$ for (vi)).

In general, \ourModelOne starkly outperforms all baselines tested %
(Table
\ref{tab:coverage}).  %
By generating $300$ peaks,
\ourModelOne is able to cover on average  $91\%$ and $72\%$ of the true formulae in the
ground truth test set for \nistData and \gnpsData respectively. Our  difference- and forward-only directional prediction ablations demonstrate the benefits of modeling both the
atom counts for each element and the differences in counts from the original molecule.

\begin{table}[t]
\centering
\caption{Model coverage (higher better) of true peak formulae as determined by \MAGMA at various max formula cutoffs for the \nistData and \gnpsData datasets. Best result for each column is in bold. Results are computed for a single test set; all re-trained models (i.e., Autoregressive and SCARF variants) are averaged across three random seeds.}
\label{tab:coverage}
\begin{tabular}{lllllllll}
\toprule
Dataset& \multicolumn{4}{l}{\nistData} & \multicolumn{4}{l}{\gnpsData}\\
\cmidrule(r){2-5} \cmidrule(r){6-9}
Coverage @ &     10 &     30 &     300 &  1000 & 10 &     30 &     300 &  1000 \\
\midrule
Random    &  0.009 & 0.026 &  0.232 &  0.532 & 0.004 &    0.014 &    0.126 &  0.336 \\
Frequency&  0.173 & 0.275 &  0.659 &  0.830 &  0.090 &    0.151 &    0.466 &  0.688 \\
CFM-ID    &  0.197 & 0.282 &     -- &     -- &  \textbf{0.170} &    0.267 &       -- &     -- \\
Autoregressive    &  0.204 & 0.262 &     0.309 &    0.317 &  0.072 &    0.082 &      0.095 &     0.099 \\
\midrule
\midrule
\ourModelOne-D   &  0.248 & 0.425 &  0.839 &  0.941 &  0.158 &    0.284 &    0.681 &  0.856 \\
\ourModelOne-F   &  0.249 & 0.476 &  0.855 &  0.943 &  0.155 &    0.306 &    0.708 &  0.859 \\
\ourModelOne      &  \textbf{0.308} & \textbf{0.552} &  \textbf{0.907} &  \textbf{0.968} &  0.164 &    \textbf{0.309} &    \textbf{0.724} &  \textbf{0.879} \\
\bottomrule
\end{tabular}
\end{table}

\paragraph{Predicting mass spectra.}
We next evaluate the strength of \ourModelTwo for intensity
prediction on the same test dataset. We compare against five baselines: a fragmentation-based approach, \cfmModel \citep{allen_competitive_2015}; two
NEIMS binned prediction models (\citep{wei_rapid_2019}; \S \ref{sec:neims}), using either feed forward network modules (FFNs), as
in the original work, or graph neural network modules (GNNs) as in \ourModelTwo and described by \citet{zhu_using_2020}; a retrained variant of 3DMolMS (\citep{hong20233dmolms}; \S \ref{sec:3dmolms}), a binned spectrum predictor that utilizes a point cloud neural network over a single molecular conformer input generated by RDKit \citep{landrum2016rdkit}; and FixedVocab, a formula prediction model that predicts intensities at a fixed library of formulae and formulae differences inspired by GRAFF-MS (\citep{murphy2023efficiently}; \S \ref{sec:fixedvocab}).

To enable fair comparison across models, we
predict test spectra at 15k bins (0.1 bin resolution between $0$ and $1500$)
with a maximum of 100 peaks for each predicted molecule. With the exception of \cfmModel, all models are hyperparameter optimized (\S \ref{sec:hyps}), retrained completely, and conditioned on the same covariate inputs as \ourModel; such steps lead to large performance boosts to the prior NEIMS method in particular. We evaluate the quality
of our predictions based upon four core criteria reflecting our original
desiderata of accuracy, physical-sensibleness, and speed: 
\begin{enumerate}
    \item \textit{Cosine sim.}: Cosine similarity between the ground truth and predicted spectra, indicating spectrum prediction accuracy.
    \item \textit{Coverage}: The fraction of ground truth spectrum peaks covered by the predicted spectrum.
    \item \textit{Valid}: The fraction of predicted peaks that can be explained by a subformula (that obeys basic ring-double bond equivalent heuristics \cite{pretsch2000structure}) of the predicted molecule. 
    \item \textit{Time (s)}: The wall time it takes (using a single CPU and no batched calculations) to load the model and predict spectra for 100 randomly selected molecules.
\end{enumerate}

\ourModel is more
accurate than all other approaches on \nistData, improving cosine similarity over a GNN
binned prediction approach by over 0.02 points in \nistData and 0.01 in \gnpsData (Table \ref{tab:spec_acc}).
Further, our method is more physically grounded insofar as all predicted peaks are
guaranteed to be valid subformulae, unlike the unconstrained binned approaches, where nearly 5\% of peak predictions cannot be explained by a valid molecular formula. Importantly, \ourModel still operates 2
orders of magnitude faster than \cfmModel (Table \ref{tab:spec_acc}). 

The heterogeneity, reduced dataset size, and increased average molecular weight (Figure \ref{fig:dataset_stats}) of \gnpsData leads to substantially worse absolute performance across all models. Interestingly, in this setting, the FixedVocab approach~\cite{murphy2023efficiently} performs better, perhaps because the strict priors of formula constraints are more helpful with fewer and more challenging training examples. We further stratify results by molecule size in Figure \ref{fig:si_stratified_results}, showing that all models are generally more accurate on smaller compounds. We additionally validate that cosine similarity is not merely measuring a model's ability to predict the parent mass peak by computing a modified cosine similarity with the original molecule's mass masked (Table \ref{tab:si_spec_acc_pep}).

\begin{table}
\centering
\caption{Spectra prediction in terms of cosine similarity, coverage (proportion of ground-truth peaks that are covered by the top 100 non-zero predictions),  validity (the fraction of predicted peaks for which a chemically plausible explanation is possible), and time. Best value in each column is typeset in bold (higher is better for all metrics but time). Results are averaged across 3 random seeds on a single data split for all retrainable models (i.e,. not CFM-ID).}
\label{tab:spec_acc}
\begin{tabular}{lllllllr}
\toprule
Dataset& \multicolumn{3}{l}{\nistData} & \multicolumn{3}{l}{\gnpsData} & \\
\cmidrule(r){2-4} \cmidrule(r){5-7}
{Metric} &  Cosine sim. &  Coverage &  Valid & Cosine sim. &  Coverage &  Valid &  Time (s)\\
\midrule
CFM-ID      &       0.412 &    0.278 &  \textbf{1.000} &                0.377 &    0.235 &  \textbf{1.000} &  1114.7 \\
3DMolMS     &       0.510 &    0.734 &  0.945 &                0.394 &    0.507 &  0.919 &     \textbf{3.5} \\
FixedVocab  &       0.704 &    0.788 &  0.997 &                \textbf{0.568} &    \textbf{0.563} &  0.998 &     5.5 \\
NEIMS (FFN) &       0.617 &    0.746 &  0.948 &                0.491 &    0.524 &  0.949 &     3.9 \\
NEIMS (GNN) &       0.694 &    0.780 &  0.947 &                0.521 &    0.547 &  0.943 &     4.9 \\
\midrule
\midrule
\ourModel{}       &       \textbf{0.726} &    \textbf{0.807} &  \textbf{1.000} &                0.536 &    0.552 &  \textbf{1.000} &    21.1 \\\bottomrule
\end{tabular}
\end{table}

\subsection{Retrieval}
\label{sec:retrieval}

A key application for forward spectrum prediction is to use predicted
spectra to determine the most plausible molecular structure assignment. We posit forward spectrum prediction models should be particularly helpful in
differentiating structurally similar molecules and design a retrieval task to
showcase such potential. 
For each test set molecule, we extract 49 potential ``decoy" options based upon
the most structurally similar \emph{isomers} (i.e., compounds with
the same precursor formula) within PubChem \cite{kim_pubchem_2016} as judged by Tanimoto similarity using Morgan fingerprints. While retrieval could be conducted on the entirety of PubChem or other similarly large molecular databases, we believe this subset retrieval setting is more practical and better mirrors a real-world setting (see \S\ref{sec:pubchem_justif} for justification).
We predict the spectra
for all molecules and rank them according to their similarity to the ground truth spectrum, computing the \emph{accuracy} for retrieval. Herein, we specifically emphasize models and retrieval on the \nistData dataset, as it is a much larger and higher quality dataset.

\begin{figure}[t]
  \centering
  \includegraphics[width=\textwidth]{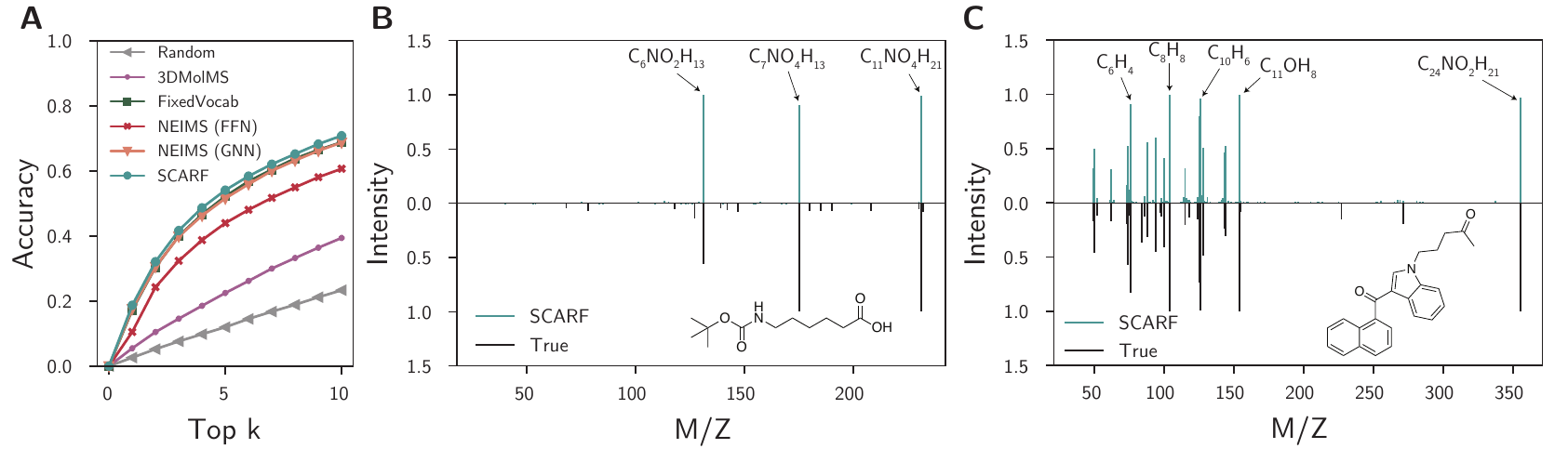}
  \caption{\ourModel enables more accurate retrieval of ground truth molecules within the \nistData dataset. \textbf{A.} Average retrieval accuracy of \ourModel at various top k thresholds. Retrieval is conducted on the same test split, and retrieval accuracy is averaged across models trained for three separate random seeds. \textbf{B-C.} Example spectrum predictions made by \ourModel (top) compared to the ground truth spectrum (bottom). Up to 5 predicted peaks with the highest intensity are annotated with their molecular formula explanation as predicted by \ourModel. The full molecule is shown inset. Further examples are in the Appendix (Figure~\ref{fig:example_mols}).}
  \label{fig:retrieval}
\end{figure}

\ourModel reaches a top 1 and top 5 retrieval accuracy in this task of   18.7\%  and 54.1\% respectively, representing an improvement over the methods with the second best top 1 accuracy of 17.5\% (NEIMS (GNN)) and top 5 accuracy of 52.2\% (Fixed Vocab) (Figure \ref{fig:retrieval}A). We highlight two example predictions from \ourModel (Figure \ref{fig:retrieval}B-C), with additional randomly sampled test set predictions shown in Figure \ref{sec:extended_results}. We repeat similar experiments within \gnpsData, but find that cosine similarity performance is uncorrelated with relative ranking performance; feed forward fingerprint based approaches are better at retrieval, despite relatively weak cosine similarity (\S \ref{sec:extended_results}). FixedVocab \cite{murphy2023efficiently} performs especially well on  \gnpsData, again likely due to the helpful biases imparted by constraining the formula vocabulary.  %

This result underscores previous observations regarding how database and model biases can skew retrieval results under certain settings \cite{hoffmann2023mad}. That is, models may be more or less robust for certain classes of molecules, so the composition of these classes in the retrieval library may affect the retrieval accuracy accordingly. The observed discrepancy between cosine similarity and retrieval performance can further be explained by the dataset shift required for computing retrieval accuracy; cosine similarity is evaluated on ``in-domain'' data, whereas retrieval relies also on accuracy on unlabeled data that may be ``out-of-domain.''

\section{Related work}
\label{sec:related}

\paragraph{Forward vs backward models.} Computational tools to identify mass spectra are often divided into two categories: forward and backward models.
Forward models, i.e., spectrum predictors, such as \ourModel or the methods discussed in Section~\ref{sec:background}, operate in the causal direction and try to predict the spectrum given the molecule.
Backward models start from the spectrum and predict  features or even full molecule structures. Early backward models used heuristics, expert rules, and even neural networks \citep{curry_msnet_1990, buchanan_heuristic_1969, sutherland_dendral-computer_1967}. Such approaches have more recently been augmented with kernel methods and more modern, deep representation learning techniques \citep{fan_metfid_2020, duhrkop_searching_2015, voronov2022ms2prop, Goldman2023annotating}. These models are complementary to spectrum predictors.

\paragraph{Mass spectra for proteomics.}
Although this paper has focused on small molecules, 
similar trends of deep representation learning for mass spectra are also emerging in the adjacent field of proteomics  \cite{yilmaz_novo_2022},
with \citet{Shouman2022-pi} recently proposing a benchmark challenge in this domain.
While small molecule and protein spectra are similar, proteomics spectra tend to be more easily predicted as fragments are often formed at peptide bonds.  We believe adapting \ourModel to this task would be an interesting direction for future work.

\paragraph{Neural set generation.}
Our work is also related to methods for modeling sets and multisets.
\ourModelOne generates a set as output, which has been studied elsewhere in the context of n-gram generation \citep{Vinyals2016-zz},
object detection \citep{Zhang2019-ld, Locatello2020-tw}, and point cloud generation \citep{Kosiorek2020-ci}.
The product formulae sets we generate, however, are different to those considered in these other task; in our setting, each member of the set (i.e., individual product formula) represents a multiset of atom types (i.e., multiple carbons, multiple hydrogens, etc.) and is constrained physically by the precursor formula.

\section{Conclusion}
\label{sec:conclusions}

In this paper we introduced \ourModel, an approach utilizing prefix tree data structures to efficiently decode mass spectra from molecules.  By first predicting product formulae and then assigning these formulae
intensities, we are able to combine the advantages of previous neural and
fragment based approaches, providing fast and physically grounded predictions. We show how these resulting predictions are both more accurate in predicting experimentally-observed spectra and yield improvements in identifying a molecule's structure from its respective mass spectrum, as tested on a widely used dataset.

In term of limitations, our model is data dependent, as indicated by the relative performance across the \nistData and \gnpsData datasets. \ourModel is also highly reliant upon the quality of product formula label assignment. The current commercial status of mass spectrometry training data poses a barrier to entry, and identifying high quality public domain data is critical for future studies. A key contribution of this work is to retrain and optimize the hyperparameters of  competing methods on a publicly available dataset under equivalent conditions to allow for future extensions. Directly training on a ranking-based loss or learning a model specific spectra distance function may be one way to improve upon our model's performance in the retrieval setting, and we outline additional potential ideas more explicitly in \S \ref{sec:future_work}.

Future directions will involve further developing SCARF for real world use cases such as unknown metabolite elucidation in clinical samples. Specific directions will include more carefully modeling covariates (e.g.,
collision energies and MS/MS instrument types), grounding product formulae in molecular graph substructures, and
utilizing such models to augment inverse spectrum-to-molecule annotation tools.

\section*{Acknowledgements}
We thank members of the Coley Research Group, as well as Michael Murphy, for
helpful discussions and comments. S.G. thanks the MIT-Takeda Program for financial support.  J.B., J.X., and C.W.C. thank the Machine Learning for Pharmaceutical Discovery and Synthesis consortium for additional support.

{
\bibliography{main}

\newcommand{\iclrConference}{International Conference on Learning
  Representations}\newcommand{\neuripsConference}[1]{Advances in Neural
  Information Processing Systems #1}\newcommand{\icmlConference}[1]{Proceedings
  of the {#1} International Conference on Machine Learning}
\begin{thebibliography}{59}
\providecommand{\natexlab}[1]{#1}
\providecommand{\url}[1]{\texttt{#1}}
\expandafter\ifx\csname urlstyle\endcsname\relax
  \providecommand{\doi}[1]{doi: #1}\else
  \providecommand{\doi}{doi: \begingroup \urlstyle{rm}\Url}\fi

\bibitem[Akiba et~al.(2019)Akiba, Sano, Yanase, Ohta, and
  Koyama]{akiba2019optuna}
Takuya Akiba, Shotaro Sano, Toshihiko Yanase, Takeru Ohta, and Masanori Koyama.
\newblock {Optuna: A Next-generation Hyperparameter Optimization Framework}.
\newblock In \emph{Proceedings of the 25th ACM SIGKDD international conference
  on knowledge discovery \& data mining}, pages 2623--2631, 2019.

\bibitem[Allen et~al.(2015)Allen, Greiner, and Wishart]{allen_competitive_2015}
Felicity Allen, Russ Greiner, and David Wishart.
\newblock {Competitive fragmentation modeling of {ESI}-{MS}/{MS} spectra for
  putative metabolite identification}.
\newblock \emph{Metabolomics}, 11\penalty0 (1):\penalty0 98--110, 2015.

\bibitem[Battaglia et~al.(2018)Battaglia, Hamrick, Bapst, Sanchez-Gonzalez,
  Zambaldi, Malinowski, Tacchetti, Raposo, Santoro, Faulkner, Gulcehre, Song,
  Ballard, Gilmer, Dahl, Vaswani, Allen, Nash, Langston, Dyer, Heess, Wierstra,
  Kohli, Botvinick, Vinyals, Li, and Pascanu]{Battaglia2018-im}
Peter~W Battaglia, Jessica~B Hamrick, Victor Bapst, Alvaro Sanchez-Gonzalez,
  Vinicius Zambaldi, Mateusz Malinowski, Andrea Tacchetti, David Raposo, Adam
  Santoro, Ryan Faulkner, Caglar Gulcehre, Francis Song, Andrew Ballard, Justin
  Gilmer, George Dahl, Ashish Vaswani, Kelsey Allen, Charles Nash, Victoria
  Langston, Chris Dyer, Nicolas Heess, Daan Wierstra, Pushmeet Kohli, Matt
  Botvinick, Oriol Vinyals, Yujia Li, and Razvan Pascanu.
\newblock {Relational inductive biases, deep learning, and graph networks}.
\newblock \emph{arXiv preprint arXiv:1806.01261}, 2018.

\bibitem[Bauermeister et~al.(2022)Bauermeister, Mannochio-Russo, Costa-Lotufo,
  Jarmusch, and Dorrestein]{bauermeister2022mass}
Anelize Bauermeister, Helena Mannochio-Russo, Let{\'\i}cia~V Costa-Lotufo,
  Alan~K Jarmusch, and Pieter~C Dorrestein.
\newblock {Mass spectrometry-based metabolomics in microbiome investigations}.
\newblock \emph{Nature Reviews Microbiology}, 20\penalty0 (3):\penalty0
  143--160, 2022.

\bibitem[Bittremieux et~al.(2022{\natexlab{a}})Bittremieux, Schmid, Huber,
  van~der Hooft, Wang, and Dorrestein]{Bittremieux2022-nk}
Wout Bittremieux, Robin Schmid, Florian Huber, Justin J~J van~der Hooft,
  Mingxun Wang, and Pieter~C Dorrestein.
\newblock {Comparison of Cosine, Modified Cosine, and Neutral Loss Based
  Spectrum Alignment For Discovery of Structurally Related Molecules}.
\newblock \emph{Journal of the American Society for Mass Spectrometry},
  33\penalty0 (9):\penalty0 1733--1744, 2022{\natexlab{a}}.

\bibitem[Bittremieux et~al.(2022{\natexlab{b}})Bittremieux, Wang, and
  Dorrestein]{bittremieux2022critical}
Wout Bittremieux, Mingxun Wang, and Pieter~C Dorrestein.
\newblock {The critical role that spectral libraries play in capturing the
  metabolomics community knowledge}.
\newblock \emph{Metabolomics}, 18\penalty0 (12):\penalty0 94,
  2022{\natexlab{b}}.

\bibitem[Bronstein et~al.(2021)Bronstein, Bruna, Cohen, and Veli{\v
  c}kovi{\'c}]{Bronstein2021-dx}
Michael~M Bronstein, Joan Bruna, Taco Cohen, and Petar Veli{\v c}kovi{\'c}.
\newblock {Geometric Deep Learning: Grids, Groups, Graphs, Geodesics, and
  Gauges}.
\newblock \emph{arXiv preprint arXiv:2104.13478}, 2021.

\bibitem[Buchanan et~al.(1969)Buchanan, Sutherland, and
  Feigenbaum]{buchanan_heuristic_1969}
Bruce Buchanan, Georgia Sutherland, and Edward~A. Feigenbaum.
\newblock {Heuristic {Dendral}: {A} program for generating explanatory
  hypotheses}.
\newblock \emph{Machine Intelligence}, 4:\penalty0 209--254, 1969.

\bibitem[Chen et~al.(2001)Chen, Chen, Jiang, Fu, Xin, and
  Zhao]{chen2001rearrangement}
Jing Chen, Yi~Chen, Yang Jiang, Hua Fu, Bin Xin, and Yu-Fen Zhao.
\newblock {Rearrangement of {P-N} to {P-O} bonds in mass spectra of
  {N-diisopropyloxyphosphoryl} amino acids/alcohols}.
\newblock \emph{Rapid Communications in Mass Spectrometry}, 15\penalty0
  (20):\penalty0 1936--1940, 2001.

\bibitem[Curry and Rumelhart(1990)]{curry_msnet_1990}
Bo~Curry and David~E. Rumelhart.
\newblock {{MSnet}: {A} neural network which classifies mass spectra}.
\newblock \emph{Tetrahedron Computer Methodology}, 3\penalty0 (3-4):\penalty0
  213--237, 1990.

\bibitem[Demarque et~al.(2016)Demarque, Crotti, Vessecchi, Lopes, and
  Lopes]{demarque2016fragmentation}
Daniel~P Demarque, Antonio~EM Crotti, Ricardo Vessecchi, Jo{\~a}o~LC Lopes, and
  Norberto~P Lopes.
\newblock {Fragmentation reactions using electrospray ionization mass
  spectrometry: an important tool for the structural elucidation and
  characterization of synthetic and natural products}.
\newblock \emph{Natural Product Reports}, 33\penalty0 (3):\penalty0 432--455,
  2016.

\bibitem[D{\"u}hrkop et~al.(2015)D{\"u}hrkop, Shen, Meusel, Rousu, and
  B{\"o}cker]{duhrkop_searching_2015}
Kai D{\"u}hrkop, Huibin Shen, Marvin Meusel, Juho Rousu, and Sebastian
  B{\"o}cker.
\newblock {Searching molecular structure databases with tandem mass spectra
  using {CSI}:{FingerID}}.
\newblock \emph{Proceedings of the National Academy of Sciences}, 112\penalty0
  (41):\penalty0 12580--12585, 2015.

\bibitem[D{\"u}hrkop et~al.(2019)D{\"u}hrkop, Fleischauer, Ludwig, Aksenov,
  Melnik, Meusel, Dorrestein, Rousu, and B{\"o}cker]{duhrkop_sirius_2019}
Kai D{\"u}hrkop, Markus Fleischauer, Marcus Ludwig, Alexander~A. Aksenov,
  Alexey~V. Melnik, Marvin Meusel, Pieter~C. Dorrestein, Juho Rousu, and
  Sebastian B{\"o}cker.
\newblock {{SIRIUS} 4: a rapid tool for turning tandem mass spectra into
  metabolite structure information}.
\newblock \emph{Nature Methods}, 16\penalty0 (4):\penalty0 299--302, 2019.

\bibitem[D{\"u}hrkop et~al.(2021)D{\"u}hrkop, Nothias, Fleischauer, Reher,
  Ludwig, Hoffmann, Petras, Gerwick, Rousu, and
  Dorrestein]{duhrkop_systematic_2021}
Kai D{\"u}hrkop, Louis-F{\'e}lix Nothias, Markus Fleischauer, Raphael Reher,
  Marcus Ludwig, Martin~A. Hoffmann, Daniel Petras, William~H. Gerwick, Juho
  Rousu, and Pieter~C. Dorrestein.
\newblock {Systematic classification of unknown metabolites using
  high-resolution fragmentation mass spectra}.
\newblock \emph{Nature Biotechnology}, 39\penalty0 (4):\penalty0 462--471,
  2021.

\bibitem[Falcon and {The PyTorch Lightning
  team}(2019)]{Falcon_PyTorch_Lightning_2019}
William Falcon and {The PyTorch Lightning team}.
\newblock {{PyTorch Lightning}}, 2019.

\bibitem[Fan et~al.(2020)Fan, Alley, Ghaffari, and Ressom]{fan_metfid_2020}
Ziling Fan, Amber Alley, Kian Ghaffari, and Habtom~W. Ressom.
\newblock {{MetFID}: artificial neural network-based compound fingerprint
  prediction for metabolite annotation}.
\newblock \emph{Metabolomics}, 16\penalty0 (10):\penalty0 104, 2020.

\bibitem[Gasteiger et~al.(1992)Gasteiger, Hanebeck, and
  Schulz]{Gasteiger1992-aj}
J~Gasteiger, W~Hanebeck, and K~P Schulz.
\newblock {Prediction of Mass Spectra from Structural Information}.
\newblock \emph{Journal of Chemical Information and Computer Sciences},
  32\penalty0 (4):\penalty0 264--271, 1992.

\bibitem[Goldman et~al.(2023{\natexlab{a}})Goldman, Li, and
  Coley]{goldman_generating_2023}
Samuel Goldman, Janet Li, and Connor~W. Coley.
\newblock {Generating {Molecular} {Fragmentation} {Graphs} with
  {Autoregressive} {Neural} {Networks}}.
\newblock \emph{arXiv preprint arXiv:2304.13136}, 2023{\natexlab{a}}.

\bibitem[Goldman et~al.(2023{\natexlab{b}})Goldman, Wohlwend, Stra{\v{z}}ar,
  Haroush, Xavier, and Coley]{Goldman2023annotating}
Samuel Goldman, Jeremy Wohlwend, Martin Stra{\v{z}}ar, Guy Haroush, Ramnik~J
  Xavier, and Connor~W Coley.
\newblock {Annotating metabolite mass spectra with domain-inspired chemical
  formula transformers}.
\newblock \emph{Nature Machine Intelligence}, 5\penalty0 (9):\penalty0
  965--979, 2023{\natexlab{b}}.

\bibitem[Hamilton(2020)]{Hamilton2020-op}
William~L Hamilton.
\newblock {Graph Representation Learning}.
\newblock \emph{Synthesis Lectures on Artificial Intelligence and Machine
  Learning}, 14\penalty0 (3):\penalty0 1--159, 2020.

\bibitem[Hochreiter and Schmidhuber(1997)]{hochreiter1997long}
Sepp Hochreiter and J{\"u}rgen Schmidhuber.
\newblock {Long short-term memory}.
\newblock \emph{Neural Computation}, 9\penalty0 (8):\penalty0 1735--1780, 1997.

\bibitem[Hoffmann et~al.(2023)Hoffmann, Kretschmer, Ludwig, and
  B{\"o}cker]{hoffmann2023mad}
Martin~A Hoffmann, Fleming Kretschmer, Marcus Ludwig, and Sebastian B{\"o}cker.
\newblock {{Mad} {Hatter} correctly annotates 98\% of small molecule tandem
  mass spectra searching in {PubChem}}.
\newblock \emph{Metabolites}, 13\penalty0 (3):\penalty0 314, 2023.

\bibitem[Hong et~al.(2023)Hong, Li, Welch, Tichy, Ye, and
  Tang]{hong20233dmolms}
Yuhui Hong, Sujun Li, Christopher~J Welch, Shane Tichy, Yuzhen Ye, and Haixu
  Tang.
\newblock {3DMolMS: prediction of tandem mass spectra from 3D molecular
  conformations}.
\newblock \emph{Bioinformatics}, 39\penalty0 (6):\penalty0 btad354, 2023.

\bibitem[Huber et~al.(2020)Huber, Verhoeven, Meijer, Spreeuw, Castilla, Geng,
  j.~van~der Hooft, Rogers, Belloum, Diblen, and
  Spaaks]{huber_matchms-processing_2020}
Florian Huber, Stefan Verhoeven, Christiaan Meijer, Hanno Spreeuw, Efra{\'\i}n
  Manuel~Villanueva Castilla, Cunliang Geng, Justin~J. j.~van~der Hooft, Simon
  Rogers, Adam Belloum, Faruk Diblen, and Jurriaan~H. Spaaks.
\newblock {matchms - processing and similarity evaluation of mass spectrometry
  data}.
\newblock \emph{Journal of Open Source Software}, 5:\penalty0 2411, 2020.

\bibitem[Huber et~al.(2021)Huber, van~der Burg, van~der Hooft, and
  Ridder]{huber_ms2deepscore_2021}
Florian Huber, Sven van~der Burg, Justin~JJ van~der Hooft, and Lars Ridder.
\newblock {{MS2DeepScore}: a novel deep learning similarity measure to compare
  tandem mass spectra}.
\newblock \emph{Journal of Cheminformatics}, 13\penalty0 (1):\penalty0 1--14,
  2021.

\bibitem[Hufsky et~al.(2014)Hufsky, Scheubert, and
  B{\"o}cker]{hufsky2014computational}
Franziska Hufsky, Kerstin Scheubert, and Sebastian B{\"o}cker.
\newblock {Computational mass spectrometry for small-molecule fragmentation}.
\newblock \emph{TrAC Trends in Analytical Chemistry}, 53:\penalty0 41--48,
  2014.

\bibitem[Kim et~al.(2021)Kim, Wang, Leber, Nothias, Reher, Kang, Van Der~Hooft,
  Dorrestein, Gerwick, and Cottrell]{kim_npclassifier_2021}
Hyun~Woo Kim, Mingxun Wang, Christopher~A. Leber, Louis-F{\'e}lix Nothias,
  Raphael Reher, Kyo~Bin Kang, Justin~JJ Van Der~Hooft, Pieter~C. Dorrestein,
  William~H. Gerwick, and Garrison~W. Cottrell.
\newblock {{NPClassifier}: a deep neural network-based structural
  classification tool for natural products}.
\newblock \emph{Journal of Natural Products}, 84\penalty0 (11):\penalty0
  2795--2807, 2021.

\bibitem[Kim et~al.(2016)Kim, Thiessen, Bolton, Chen, Fu, Gindulyte, Han, He,
  He, and Shoemaker]{kim_pubchem_2016}
Sunghwan Kim, Paul~A. Thiessen, Evan~E. Bolton, Jie Chen, Gang Fu, Asta
  Gindulyte, Lianyi Han, Jane He, Siqian He, and Benjamin~A. Shoemaker.
\newblock {PubChem Substance and Compound databases}.
\newblock \emph{Nucleic Acids Research}, 44\penalty0 (D1):\penalty0
  D1202--D1213, 2016.

\bibitem[Kosiorek et~al.(2020)Kosiorek, Kim, and Rezende]{Kosiorek2020-ci}
Adam~R Kosiorek, Hyunjik Kim, and Danilo~J Rezende.
\newblock {Conditional Set Generation with Transformers}.
\newblock In \emph{Workshop on Object-Oriented Learning at ICML}, 2020.

\bibitem[Landrum(2016)]{landrum2016rdkit}
Greg Landrum.
\newblock {RDKit: Open-source cheminformatics software}, 2016.

\bibitem[Lee et~al.(2019)Lee, Lee, Kim, Kosiorek, Choi, and Teh]{lee2019set}
Juho Lee, Yoonho Lee, Jungtaek Kim, Adam Kosiorek, Seungjin Choi, and Yee~Whye
  Teh.
\newblock {Set Transformer: A Framework for Attention-based
  Permutation-Invariant Neural Networks}.
\newblock In \emph{\icmlConference{36th}}, pages 3744--3753, 2019.

\bibitem[Li et~al.(2016)Li, Tarlow, Brockschmidt, and Zemel]{Li2015-fc}
Yujia Li, Daniel Tarlow, Marc Brockschmidt, and Richard Zemel.
\newblock {Gated Graph Sequence Neural Networks}.
\newblock In \emph{\iclrConference{}}, 2016.

\bibitem[Liaw et~al.(2018)Liaw, Liang, Nishihara, Moritz, Gonzalez, and
  Stoica]{liaw2018tune}
Richard Liaw, Eric Liang, Robert Nishihara, Philipp Moritz, Joseph~E Gonzalez,
  and Ion Stoica.
\newblock {Tune: A Research Platform for Distributed Model Selection and
  Training}.
\newblock In \emph{ICML AutoML Workshop}, 2018.

\bibitem[Locatello et~al.(2020)Locatello, Weissenborn, Unterthiner, Mahendran,
  Heigold, Uszkoreit, Dosovitskiy, and Kipf]{Locatello2020-tw}
Francesco Locatello, Dirk Weissenborn, Thomas Unterthiner, Aravindh Mahendran,
  Georg Heigold, Jakob Uszkoreit, Alexey Dosovitskiy, and Thomas Kipf.
\newblock {Object-centric learning with Slot Attention}.
\newblock In \emph{\neuripsConference{33}}, 2020.

\bibitem[Ludwig et~al.(2020)Ludwig, Nothias, D{\"u}hrkop, Koester, Fleischauer,
  Hoffmann, Petras, Vargas, Morsy, and
  Aluwihare]{ludwig_database-independent_2020}
Marcus Ludwig, Louis-F{\'e}lix Nothias, Kai D{\"u}hrkop, Irina Koester, Markus
  Fleischauer, Martin~A. Hoffmann, Daniel Petras, Fernando Vargas, Mustafa
  Morsy, and Lihini Aluwihare.
\newblock {Database-independent molecular formula annotation using {Gibbs}
  sampling through {ZODIAC}}.
\newblock \emph{Nature Machine Intelligence}, 2\penalty0 (10):\penalty0
  629--641, 2020.

\bibitem[Murphy et~al.(2023)Murphy, Jegelka, Fraenkel, Kind, Healey, and
  Butler]{murphy2023efficiently}
Michael Murphy, Stefanie Jegelka, Ernest Fraenkel, Tobias Kind, David Healey,
  and Thomas Butler.
\newblock {Efficiently predicting high resolution mass spectra with graph
  neural networks}.
\newblock In \emph{\icmlConference{40th}}, 2023.

\bibitem[NIST(2020)]{noauthor_tandem_nodate}
NIST.
\newblock {Tandem {Mass} {Spectral} {Library}}, 2020.

\bibitem[Paszke et~al.(2019)Paszke, Gross, Massa, Lerer, Bradbury, Chanan,
  Killeen, Lin, Gimelshein, Antiga, et~al.]{paszke2019pytorch}
Adam Paszke, Sam Gross, Francisco Massa, Adam Lerer, James Bradbury, Gregory
  Chanan, Trevor Killeen, Zeming Lin, Natalia Gimelshein, Luca Antiga, et~al.
\newblock {PyTorch: An Imperative Style, High-Performance Deep Learning
  Library}.
\newblock In \emph{\neuripsConference{32}}, 2019.

\bibitem[Pretsch et~al.(2000)Pretsch, B{\"u}hlmann, Affolter, Pretsch,
  Bhuhlmann, and Affolter]{pretsch2000structure}
Ern{\"o} Pretsch, Philippe B{\"u}hlmann, Christian Affolter, Ernho Pretsch,
  P~Bhuhlmann, and C~Affolter.
\newblock \emph{{Structure determination of organic compounds}}.
\newblock Springer, 2000.

\bibitem[Quinn et~al.(2020)Quinn, Melnik, Vrbanac, Fu, Patras, Christy, Bodai,
  Belda-Ferre, Tripathi, Chung, Downes, Welch, Quinn, Humphrey, Panitchpakdi,
  Weldon, Aksenov, da~Silva, Avila-Pacheco, Clish, Bae, Mallick, Franzosa,
  Lloyd-Price, Bussell, Thron, Nelson, Wang, Leszczynski, Vargas, Gauglitz,
  Meehan, Gentry, Arthur, Komor, Poulsen, Boland, Chang, Sandborn, Lim, Garg,
  Lumeng, Xavier, Kazmierczak, Jain, Egan, Rhee, Ferguson, Raffatellu,
  Vlamakis, Haddad, Siegel, Huttenhower, Mazmanian, Evans, Nizet, Knight, and
  Dorrestein]{quinn_global_2020}
Robert~A. Quinn, Alexey~V. Melnik, Alison Vrbanac, Ting Fu, Kathryn~A. Patras,
  Mitchell~P. Christy, Zsolt Bodai, Pedro Belda-Ferre, Anupriya Tripathi,
  Lawton~K. Chung, Michael Downes, Ryan~D. Welch, Melissa Quinn, Greg Humphrey,
  Morgan Panitchpakdi, Kelly~C. Weldon, Alexander Aksenov, Ricardo da~Silva,
  Julian Avila-Pacheco, Clary Clish, Sena Bae, Himel Mallick, Eric~A. Franzosa,
  Jason Lloyd-Price, Robert Bussell, Taren Thron, Andrew~T. Nelson, Mingxun
  Wang, Eric Leszczynski, Fernando Vargas, Julia~M. Gauglitz, Michael~J.
  Meehan, Emily Gentry, Timothy~D. Arthur, Alexis~C. Komor, Orit Poulsen,
  Brigid~S. Boland, John~T. Chang, William~J. Sandborn, Meerana Lim, Neha Garg,
  Julie~C. Lumeng, Ramnik~J. Xavier, Barbara~I. Kazmierczak, Ruchi Jain, Marie
  Egan, Kyung~E. Rhee, David Ferguson, Manuela Raffatellu, Hera Vlamakis,
  Gabriel~G. Haddad, Dionicio Siegel, Curtis Huttenhower, Sarkis~K. Mazmanian,
  Ronald~M. Evans, Victor Nizet, Rob Knight, and Pieter~C. Dorrestein.
\newblock {Global chemical effects of the microbiome include new bile-acid
  conjugations}.
\newblock \emph{Nature}, 579\penalty0 (7797):\penalty0 123--129, 2020.

\bibitem[Ridder et~al.(2014)Ridder, van~der Hooft, and
  Verhoeven]{ridder_automatic_2014}
Lars Ridder, Justin~JJ van~der Hooft, and Stefan Verhoeven.
\newblock {Automatic compound annotation from mass spectrometry data using
  {MAGMa}}.
\newblock \emph{Mass Spectrometry}, 3\penalty0 (Spec Iss 2):\penalty0 S0033,
  2014.

\bibitem[Shouman et~al.(2022)Shouman, Gabriel, Giurcoiu, Sternlicht, and
  Wilhelm]{Shouman2022-pi}
Omar Shouman, Wassim Gabriel, Victor-George Giurcoiu, Vitor Sternlicht, and
  Mathias Wilhelm.
\newblock {{PROSPECT}: Labeled Tandem Mass Spectrometry Dataset for Machine
  Learning in Proteomics}.
\newblock In \emph{\neuripsConference{35}}, 2022.

\bibitem[Sutherland(1967)]{sutherland_dendral-computer_1967}
Georgia Sutherland.
\newblock {{Dendral}-a computer program for generating and filtering chemical
  structures}.
\newblock Technical report, Stanford University, Department of Computer
  Science, 1967.

\bibitem[Tancik et~al.(2020)Tancik, Srinivasan, Mildenhall, Fridovich-Keil,
  Raghavan, Singhal, Ramamoorthi, Barron, and Ng]{Tancik2020-yy}
Matthew Tancik, Pratul~P Srinivasan, Ben Mildenhall, Sara Fridovich-Keil,
  Nithin Raghavan, Utkarsh Singhal, Ravi Ramamoorthi, Jonathan~T Barron, and
  Ren Ng.
\newblock {Fourier Features Let Networks Learn High Frequency Functions in Low
  Dimensional Domains}.
\newblock In \emph{\neuripsConference{33}}, pages 7537--7547, 2020.

\bibitem[Vaswani et~al.(2017)Vaswani, Shazeer, Parmar, Uszkoreit, Jones, Gomez,
  Kaiser, and Polosukhin]{Vaswani2017-se}
Ashish Vaswani, Noam Shazeer, Niki Parmar, Jakob Uszkoreit, Llion Jones,
  Aidan~N Gomez, Lukasz Kaiser, and Illia Polosukhin.
\newblock {Attention Is All You Need}.
\newblock In \emph{\neuripsConference{30}}, pages 5998--6008, 2017.

\bibitem[Vinyals et~al.(2016)Vinyals, Bengio, and Kudlur]{Vinyals2016-zz}
Oriol Vinyals, Samy Bengio, and Manjunath Kudlur.
\newblock {Order Matters: Sequence to sequence for sets}.
\newblock In \emph{\iclrConference}, 2016.

\bibitem[Voronov et~al.(2022)Voronov, Frandsen, Bargh, Healey, Lightheart,
  Kind, Dorrestein, Colluru, and Butler]{voronov2022ms2prop}
Gennady Voronov, Abe Frandsen, Brian Bargh, David Healey, Rose Lightheart,
  Tobias Kind, Pieter Dorrestein, Viswa Colluru, and Thomas Butler.
\newblock {{MS2Prop}: A machine learning model that directly predicts chemical
  properties from mass spectrometry data for novel compounds}.
\newblock \emph{BioRxiv}, 2022.

\bibitem[Wang et~al.(2016)Wang, Carver, Phelan, Sanchez, Garg, Peng, Nguyen,
  Watrous, Kapono, and Luzzatto-Knaan]{wang_sharing_2016}
Mingxun Wang, Jeremy~J. Carver, Vanessa~V. Phelan, Laura~M. Sanchez, Neha Garg,
  Yao Peng, Don~Duy Nguyen, Jeramie Watrous, Clifford~A. Kapono, and Tal
  Luzzatto-Knaan.
\newblock {Sharing and community curation of mass spectrometry data with
  {Global} {Natural} {Products} {Social} {Molecular} {Networking}}.
\newblock \emph{Nature Biotechnology}, 34\penalty0 (8):\penalty0 828--837,
  2016.

\bibitem[Wang et~al.(2019)Wang, Zheng, Ye, Gan, Li, Song, Zhou, Ma, Yu, Gai,
  Xiao, He, Karypis, Li, and Zhang]{Wang2019-ga}
Minjie Wang, Da~Zheng, Zihao Ye, Quan Gan, Mufei Li, Xiang Song, Jinjing Zhou,
  Chao Ma, Lingfan Yu, Yu~Gai, Tianjun Xiao, Tong He, George Karypis, Jinyang
  Li, and Zheng Zhang.
\newblock {Deep Graph Library: A Graph-Centric, Highly-Performant Package for
  Graph Neural Networks}.
\newblock \emph{arXiv preprint arXiv:1909.01315}, 2019.

\bibitem[Wei et~al.(2019)Wei, Belanger, Adams, and Sculley]{wei_rapid_2019}
Jennifer~N. Wei, David Belanger, Ryan~P. Adams, and D.~Sculley.
\newblock {Rapid Prediction of Electron–Ionization Mass Spectrometry Using
  Neural Networks}.
\newblock \emph{ACS Central Science}, 5\penalty0 (4):\penalty0 700--708, 2019.

\bibitem[Wishart(2019)]{wishart_metabolomics_2019}
David~S. Wishart.
\newblock {Metabolomics for Investigating Physiological and Pathophysiological
  Processes}.
\newblock \emph{Physiological Reviews}, 99\penalty0 (4):\penalty0 1819--1875,
  2019.

\bibitem[Wolf et~al.(2010)Wolf, Schmidt, M{\"u}ller-Hannemann, and
  Neumann]{wolf_silico_2010}
Sebastian Wolf, Stephan Schmidt, Matthias M{\"u}ller-Hannemann, and Steffen
  Neumann.
\newblock {In silico fragmentation for computer assisted identification of
  metabolite mass spectra}.
\newblock \emph{BMC Bioinformatics}, 11\penalty0 (1):\penalty0 148, 2010.

\bibitem[Xing et~al.(2023)Xing, Shen, Xu, Li, and Huan]{xing2023buddy}
Shipei Xing, Sam Shen, Banghua Xu, Xiaoxiao Li, and Tao Huan.
\newblock {BUDDY: molecular formula discovery via bottom-up MS/MS
  interrogation}.
\newblock \emph{Nature Methods}, 20:\penalty0 881--890, 2023.

\bibitem[Yilmaz et~al.(2022)Yilmaz, Fondrie, Bittremieux, Oh, and
  Noble]{yilmaz_novo_2022}
Melih Yilmaz, William Fondrie, Wout Bittremieux, Sewoong Oh, and William~S.
  Noble.
\newblock {De novo mass spectrometry peptide sequencing with a transformer
  model}.
\newblock In \emph{\icmlConference{39th}}, pages 25514--25522, 2022.

\bibitem[Young et~al.(2021)Young, Wang, and R{\"o}st]{young_massformer_2021}
Adamo Young, Bo~Wang, and Hannes R{\"o}st.
\newblock {{MassFormer}: {Tandem} {Mass} {Spectrum} {Prediction} with {Graph}
  {Transformers}}.
\newblock \emph{arXiv preprint arXiv:2111.04824}, 2021.

\bibitem[Zaheer et~al.(2017)Zaheer, Kottur, Ravanbakhsh, Poczos, Salakhutdinov,
  and Smola]{Zaheer2017-jj}
Manzil Zaheer, Satwik Kottur, Siamak Ravanbakhsh, Barnabas Poczos, Ruslan
  Salakhutdinov, and Alexander Smola.
\newblock {Deep Sets}.
\newblock In \emph{\neuripsConference{30}}, pages 3391--3401, 2017.

\bibitem[Zhang et~al.(2019)Zhang, Hare, and Prugel-Bennett]{Zhang2019-ld}
Yan Zhang, Jonathon Hare, and Adam Prugel-Bennett.
\newblock {Deep Set Prediction Networks}.
\newblock In \emph{\neuripsConference{32}}, 2019.

\bibitem[Zhu et~al.(2020)Zhu, Liu, and Hassoun]{zhu_using_2020}
Hao Zhu, Liping Liu, and Soha Hassoun.
\newblock {Using {Graph} {Neural} {Networks} for {Mass} {Spectrometry}
  {Prediction}}.
\newblock In \emph{Machine Learning for Molecules Workshop at NeurIPS}, 2020.

\bibitem[Zhu and Jonas(2023)]{zhu2023rapid}
Richard~Licheng Zhu and Eric Jonas.
\newblock {Rapid Approximate Subset-Based Spectra Prediction for Electron
  Ionization--Mass Spectrometry}.
\newblock \emph{Analytical Chemistry}, 95\penalty0 (5):\penalty0 2653--2663,
  2023.

\end{thebibliography}
\bibliographystyle{plainnat}

\small
}

\FloatBarrier
\clearpage
\appendix
\label{appendix}

\renewcommand{\theHfigure}{A\arabic{figure}}
\renewcommand{\theHtable}{A\arabic{table}}

\renewcommand{\thetable}{\Alph{section}\arabic{table}}
\renewcommand{\theequation}{\Alph{section}.\arabic{equation}}
\renewcommand{\thefigure}{\Alph{section}\arabic{figure}}
\setcounter{table}{0} %
\setcounter{figure}{0} %
\setcounter{equation}{0} %

\section{Appendix}

\subsection{Extended results}
\label{sec:extended_results}
\begin{figure}%
  \centering
  \includegraphics[width=\textwidth]{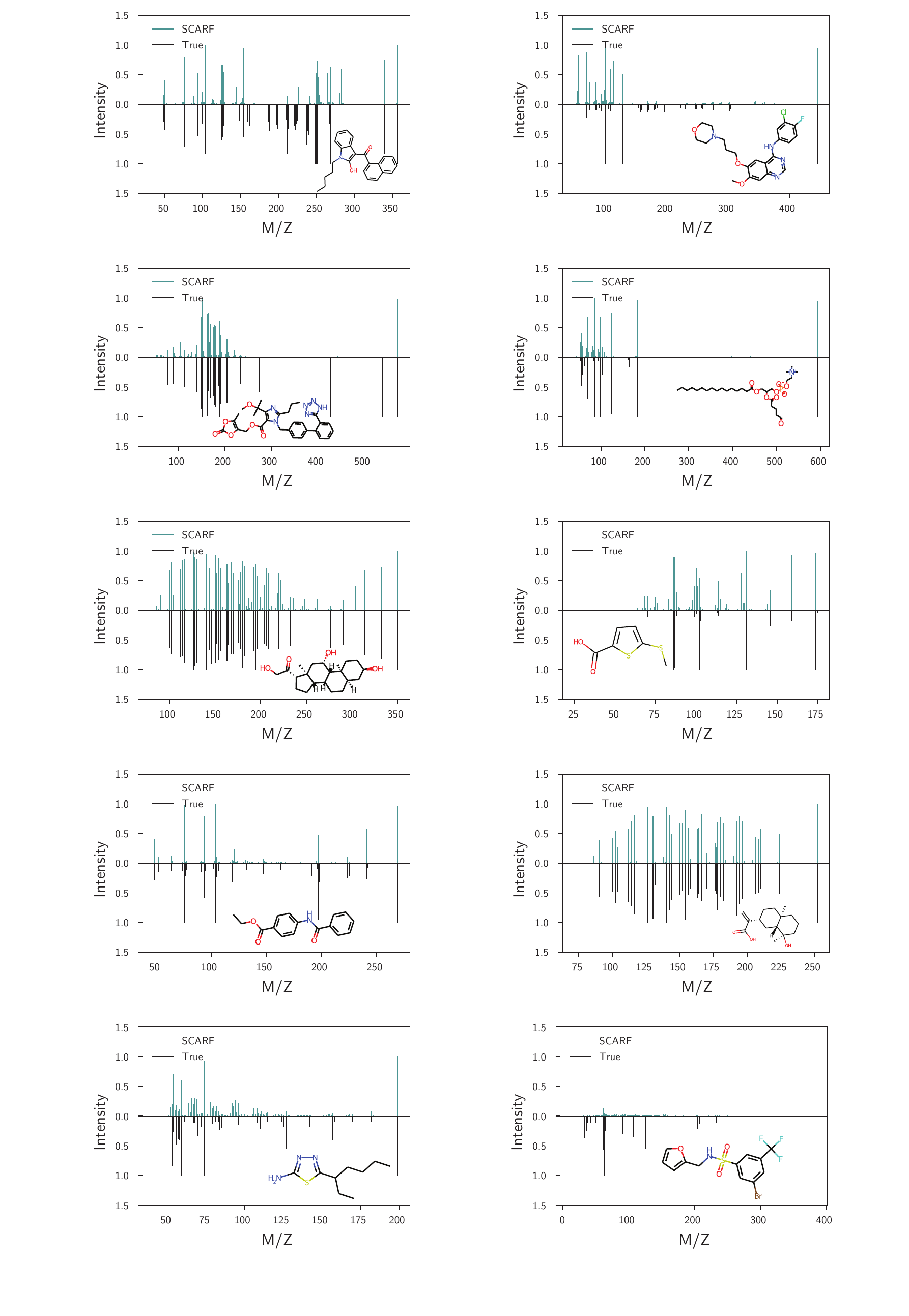}
  \caption{Example spectra predictions from the \nistData dataset for 10 randomly selected test molecules. The ground truth spectra are shown underneath in black, with predictions above in teal. Molecules are shown inset.}
  \label{fig:example_mols}

\end{figure}

\FloatBarrier
We benchmark models in terms of retrieval accuracy as described (\S~\ref{sec:retrieval}) for both the \nistData and \gnpsData datasets (Table \ref{tab:nist_spec_retrieval}, \ref{tab:canopus_spec_retrieval}). We recreate retrieval experiments using the full PubChem retrieval library in Table \ref{tab:si_retrieval_acc_all}. We reproduce main text results with standard error values included in Tables \ref{tab:coverage_gnps_app}, \ref{tab:coverage_nist_app}, and \ref{tab:spec_acc_app}. 
We showcase additional spectra predictions from our model trained on \nistData in Figure \ref{fig:example_mols}.

\begin{table}[H]
\centering
\caption{\nistData spectra prediction retrieval top k accuracy for different values of k. All values represent the mean across three separate random seeds $\pm$ the standard error of the mean for a single test set. }
\resizebox{\textwidth}{!}{
\begin{tabular}{lrrrrrrr}
\toprule
top k &     1  &     2  &     3  &     4  &     5  &     8  &       10 \\
\midrule
Random      &  $0.026 \pm 0.000$ &  $0.052 \pm 0.001$ &  $0.076 \pm 0.001$ &  $0.098 \pm 0.001$ &  $0.120 \pm 0.000$ &  $0.189 \pm 0.001$ &  $0.233 \pm 0.002$ \\
3DMolMS     &  $0.055 \pm 0.002$ &  $0.105 \pm 0.000$ &  $0.146 \pm 0.002$ &  $0.185 \pm 0.004$ &  $0.225 \pm 0.004$ &  $0.332 \pm 0.003$ &  $0.394 \pm 0.004$ \\
FixedVocab  &  $0.172 \pm 0.002$ &  $0.304 \pm 0.002$ &  $0.399 \pm 0.001$ &  $0.466 \pm 0.004$ &  $0.522 \pm 0.006$ &  $0.638 \pm 0.005$ &  $0.688 \pm 0.003$ \\
NEIMS (FFN) &  $0.105 \pm 0.002$ &  $0.243 \pm 0.006$ &  $0.324 \pm 0.006$ &  $0.387 \pm 0.006$ &  $0.440 \pm 0.007$ &  $0.549 \pm 0.005$ &  $0.607 \pm 0.002$ \\
NEIMS (GNN) &  $0.175 \pm 0.003$ &  $0.305 \pm 0.001$ &  $0.398 \pm 0.001$ &  $0.462 \pm 0.002$ &  $0.515 \pm 0.003$ &  $0.632 \pm 0.003$ &  $0.687 \pm 0.003$ \\
\midrule
\midrule
\ourModel       &  \boldmath$0.187 \pm 0.004$ &  \boldmath$0.321 \pm 0.006$ &  \boldmath$0.417 \pm 0.004$ &  \boldmath$0.486 \pm 0.004$ &  \boldmath$0.541 \pm 0.005$ &  \boldmath$0.652 \pm 0.004$ &  \boldmath$0.708 \pm 0.005$ \\
\bottomrule
\end{tabular}}
\label{tab:nist_spec_retrieval}
\end{table}

\begin{table}[H]
\centering
\caption{\gnpsData spectra prediction retrieval top k accuracy for different values of k.  All values represent the mean across three separate random seeds $\pm$ the standard error of the mean for a single test set.}
\resizebox{\textwidth}{!}{
\begin{tabular}{lrrrrrrr}
\toprule
top k &     1  &     2  &     3  &     4  &     5  &       8  &      10 \\
\midrule
\midrule
Random      &  $0.033 \pm 0.001$ &  $0.061 \pm 0.005$ &  $0.092 \pm 0.003$ &  $0.118 \pm 0.003$ &  $0.141 \pm 0.006$ &  $0.216 \pm 0.006$ &  $0.258 \pm 0.006$ \\
3DMolMS     &  $0.087 \pm 0.001$ &  $0.159 \pm 0.010$ &  $0.218 \pm 0.004$ &  $0.268 \pm 0.006$ &  $0.317 \pm 0.006$ &  $0.427 \pm 0.008$ &  $0.488 \pm 0.005$ \\
FixedVocab  &  $0.193 \pm 0.003$ &  \boldmath$0.314 \pm 0.004$ &  \boldmath$0.390 \pm 0.003$ &  \boldmath$0.448 \pm 0.005$ &  \boldmath$0.492 \pm 0.001$ &  \boldmath$0.587 \pm 0.005$ &  $0.635 \pm 0.006$ \\
NEIMS (FFN) &  \boldmath$0.195 \pm 0.003$ &  $0.313 \pm 0.002$ &  $0.388 \pm 0.003$ &  $0.447 \pm 0.006$ &  $0.488 \pm 0.002$ &  $0.585 \pm 0.007$ &  $0.624 \pm 0.010$ \\
NEIMS (GNN) &  $0.174 \pm 0.007$ &  $0.285 \pm 0.004$ &  $0.362 \pm 0.002$ &  $0.422 \pm 0.001$ &  $0.471 \pm 0.002$ &  $0.586 \pm 0.007$ &  \boldmath$0.640 \pm 0.005$ \\
\midrule
\midrule
\ourModel       &  $0.135 \pm 0.007$ &  $0.242 \pm 0.001$ &  $0.320 \pm 0.001$ &  $0.389 \pm 0.004$ &  $0.444 \pm 0.002$ &  $0.569 \pm 0.001$ &  $0.630 \pm 0.008$ \\
\bottomrule
\end{tabular}}
\label{tab:canopus_spec_retrieval}
\end{table}

\begin{table}[H]
\centering
\caption{Retrieval accuracy  on \nistData for a single 500 molecule subset of the test set using a library of 49 decoys or all decoys contained in PubChem (``None''). Results were repeated for 3 random training seeds of the model and are shown $\pm$ the standard error of the mean. The top value is shown in bold.}
\resizebox{\textwidth}{!}{
\begin{tabular}{lllllll}
\toprule
PubChem limit & \multicolumn{3}{l}{50} & \multicolumn{3}{l}{None} \\
Top k &                  1 &                  2 &                  3 &                  1 &                  2 &                  3 \\
\midrule
FixedVocab  &  $0.168 \pm 0.003$ &  $0.308 \pm 0.003$ &  $0.410 \pm 0.005$ &  $0.145 \pm 0.004$ &  \boldmath$0.258 \pm 0.004$ &  $0.323 \pm 0.001$ \\
NEIMS (FFN) &  $0.102 \pm 0.002$ &  $0.237 \pm 0.003$ &  $0.315 \pm 0.004$ &  $0.087 \pm 0.003$ &  $0.183 \pm 0.008$ &  $0.236 \pm 0.007$ \\
NEIMS (GNN) &  $0.169 \pm 0.003$ &  $0.300 \pm 0.004$ &  $0.402 \pm 0.005$ &  $0.138 \pm 0.004$ &  $0.239 \pm 0.005$ &  $0.312 \pm 0.008$ \\
\midrule\midrule
\ourModel       &  \boldmath$0.204 \pm 0.009$ &  \boldmath$0.326 \pm 0.005$ &  \boldmath$0.432 \pm 0.009$ &  \boldmath$0.167 \pm 0.003$ &  \boldmath$0.258 \pm 0.001$ &  \boldmath$0.336 \pm 0.003$ \\
\bottomrule
\end{tabular}}
\label{tab:si_retrieval_acc_all}
\end{table}

\begin{table}[ht]
\centering
\caption{Model coverage of true peak formulae as determined by \MAGMA at various max formula cutoffs for the \gnpsData dataset. Results are calculated for a single held out test split, shown $\pm$ the standard error of the mean across three random seeds for all models that were retrained. The best value (i.e., highest) is typeset in bold for each column.}
\label{tab:coverage_gnps_app}
\resizebox{\textwidth}{!}{\begin{tabular}{lllll}
\toprule 
Coverage @ &                 10 &                 30 &                300 &               1000 \\
\midrule \midrule
Random         &  $0.004$ &  $0.014$ &  $0.126$ &  $0.336$ \\
Frequency      &  $0.090$ &  $0.151$ &  $0.466$ &  $0.688$ \\
CFM-ID         &  \boldmath$0.170$ &  $0.267$ &                 -- &                 -- \\
Autoregressive &  $0.072 \pm 0.001$ &  $0.082 \pm 0.002$ &  $0.095 \pm 0.001$ &  $0.099 \pm 0.000$ \\
\midrule \midrule
\ourModel-D        &  $0.158 \pm 0.001$ &  $0.284 \pm 0.003$ &  $0.681 \pm 0.002$ &  $0.856 \pm 0.002$ \\
\ourModel-F        &  $0.155 \pm 0.002$ &  $0.306 \pm 0.003$ &  $0.708 \pm 0.003$ &  $0.859 \pm 0.001$ \\
\ourModel          &  $0.164 \pm 0.009$ &  \boldmath$0.309 \pm 0.014$ & \boldmath$0.724 \pm 0.013$ &  \boldmath$0.879 \pm 0.004$ \\
\bottomrule
\end{tabular}}
\end{table}

\begin{table}[ht]
\centering
\caption{Model coverage of true peak formulae as determined by \MAGMA at various max formula cutoffs for the \nistData dataset. Results are calculated for a single held out test split, shown $\pm$ the standard error of the mean across three random seeds for all models that were retrained. The best value (i.e., highest) is typeset in bold for each column.}
\label{tab:coverage_nist_app}
\resizebox{\textwidth}{!}{
\begin{tabular}{lllll}
\toprule
Coverage @ &                 10 &                 30 &                300 &               1000 \\
\midrule\midrule
Random         &  $0.009$ &  $0.026$ &  $0.232$ &  $0.532$ \\
Frequency      &  $0.173$ &  $0.275$ &  $0.659$ &  $0.830$ \\
CFM-ID         &  $0.197$ &  $0.282$ &                 -- &                 -- \\
Autoregressive &  $0.204 \pm 0.001$ &  $0.262 \pm 0.002$ &  $0.309 \pm 0.005$ &  $0.317 \pm 0.006$ \\
\midrule \midrule
\ourModel-D        &  $0.248 \pm 0.001$ &  $0.425 \pm 0.002$ &  $0.839 \pm 0.002$ &  $0.941 \pm 0.001$ \\
\ourModel-F        &  $0.249 \pm 0.001$ &  $0.476 \pm 0.002$ &  $0.855 \pm 0.000$ &  $0.943 \pm 0.001$ \\
\ourModel          &  \boldmath$0.308 \pm 0.002$ &  \boldmath$0.552 \pm 0.001$ &  \boldmath$0.907 \pm 0.002$ &  \boldmath$0.968 \pm 0.001$ \\
\bottomrule
\end{tabular}}
\end{table}

\begin{table}[ht]
\centering
\caption{Spectra prediction in terms of cosine similarity, coverage (proportion of ground-truth peaks that are covered by the top 100 non-zero predictions),  validity (the fraction of predicted peaks for which a chemically plausible explanation is possible), and time. Best value in each column is typeset in bold (higher is better for all metrics but time).  Values are shown $\pm$ the standard error of the mean computed across three random seeds on a single test set for all models that could be retrained (i.e., not CFM-ID).}
\label{tab:spec_acc_app}
\resizebox{\textwidth}{!}{
\begin{tabular}{llllllll}
\toprule
Dataset & \multicolumn{3}{l}{\nistData} & \multicolumn{4}{l}{\gnpsData} \\
\cmidrule(r){2-4} \cmidrule(r){5-7}
{} &          Cosine sim. &           Coverage &             Valid &        Cosine sim. &           Coverage &             Valid &   Time (s) \\
\midrule\midrule
CFM-ID      &  $0.412 \pm 0.000$ &  $0.278 \pm 0.000$ & \boldmath $1.00 \pm 0.000$ &    $0.377 \pm 0.000$ &  $0.235 \pm 0.000$ &  \boldmath$1.00 \pm 0.000$ &  $1114.7$ \\
3DMolMS     &  $0.510 \pm 0.000$ &  $0.734 \pm 0.001$ &  $0.94 \pm 0.001$ &    $0.394 \pm 0.002$ &  $0.507 \pm 0.001$ &  $0.92 \pm 0.000$ &     \boldmath$3.5$ \\
FixedVocab  &  $0.704 \pm 0.000$ &  $0.788 \pm 0.000$ &  \boldmath$1.00 \pm 0.000$ &    \boldmath$0.568 \pm 0.002$ &  \boldmath$0.563 \pm 0.001$ &  \boldmath$1.00 \pm 0.000$ &     $5.5$ \\
NEIMS (FFN) &  $0.617 \pm 0.000$ &  $0.746 \pm 0.001$ &  $0.95 \pm 0.001$ &    $0.491 \pm 0.002$ &  $0.524 \pm 0.001$ &  $0.95 \pm 0.000$ &     $3.9$ \\
NEIMS (GNN) &  $0.694 \pm 0.000$ &  $0.780 \pm 0.000$ &  $0.95 \pm 0.001$ &    $0.521 \pm 0.002$ &  $0.547 \pm 0.003$ &  $0.94 \pm 0.001$ &     $4.9$ \\
\midrule\midrule
\ourModel       &  \boldmath$0.726 \pm 0.001$ &  \boldmath$0.807 \pm 0.000$ &  \boldmath$1.00 \pm 0.000$ &    $0.536 \pm 0.007$ &  $0.552 \pm 0.008$ &  \boldmath$1.00 \pm 0.000$ &    $21.1$ \\
\end{tabular}}
\end{table}

\begin{table}
\centering
\caption{Spectra prediction accuracy comparing inclusion (Cosine sim.) and exclusion (Cosine sim. (no MS1)) of the precursor mass. For all compounds, the peak at the mass of the input compound is masked in the prediction and ground truth to compute Cosine sim. (no MS1). All results represent an average on a single test set across three random seeds.}
\label{tab:si_spec_acc_pep}

\begin{tabular}{lrrrr}
\toprule
Dataset & \multicolumn{2}{l}{\nistData} & \multicolumn{2}{l}{\gnpsData} \\
{} & Cosine sim. & Cosine sim. (no MS1) &          Cosine sim. & Cosine sim. (no MS1) \\
\midrule
CFM-ID      &       0.412 &                0.289 &                0.377 &                0.326 \\
3DMolMS     &       0.510 &                0.517 &                0.394 &                0.390 \\
FixedVocab  &       0.704 &                0.637 &                \textbf{0.568} &                \textbf{0.505} \\
NEIMS (FFN) &       0.617 &                0.557 &                0.491 &                0.454 \\
NEIMS (GNN) &       0.694 &                0.620 &                0.521 &                0.477 \\
\midrule\midrule
\ourModel       &       \textbf{0.726} &                \textbf{0.663} &                0.536 &                0.466 \\
\bottomrule
\end{tabular}
\end{table}

\begin{figure}%
  \centering
  \includegraphics[width=\textwidth]{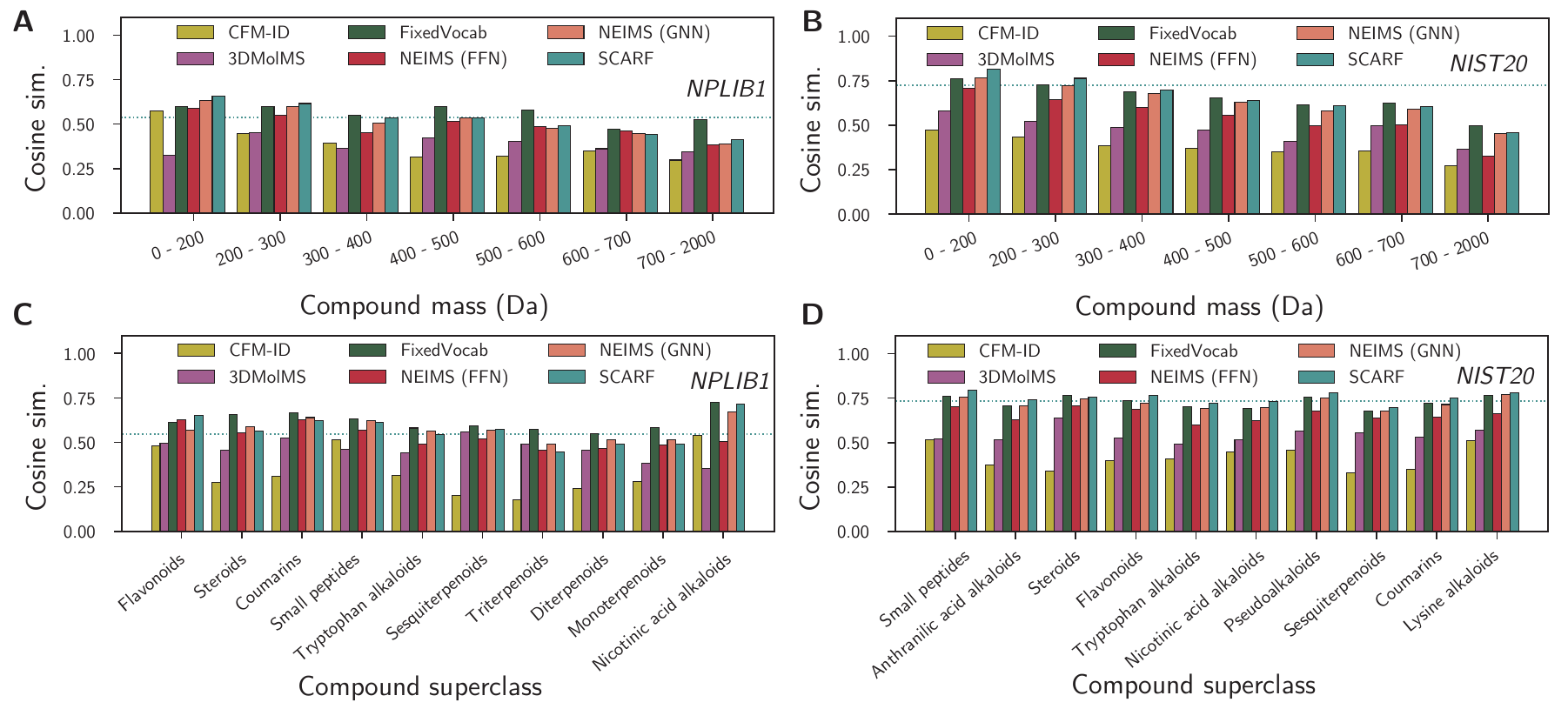}
  \caption{Cosine similarity of predicted spectra is stratified across molecular weight for both \gnpsData (\textbf{A}) and \nistData (\textbf{B}). We further stratify results across putative chemical classes of input molecules using NPClassifier \cite{kim_npclassifier_2021} for both \gnpsData (\textbf{C}) and \nistData (\textbf{D}). The dotted line indicates the average predictive cosine similarity of \ourModel across all examples and averaged over three random splits.}
  \label{fig:si_stratified_results}

\end{figure}

\begin{table}
\centering
\caption{\nistData retrieval accuracy averaged across three random seeds of model training stratified by the weight of the target molecule.}
\label{tab:si_nist20_mass_strat}
\resizebox{\textwidth}{!}{
\begin{tabular}{lrrrrrrr}
\toprule
Dataset & \multicolumn{7}{l}{\nistData} \\
Molecular weight & 0 - 200 & 200 - 300 & 300 - 400 & 400 - 500 & 500 - 600 & 600 - 700 & 700 - 2000 \\
Num. compounds     & 624 &1358& 900& 387& 129& 56 &89\\
\midrule
Random      &   0.024 &     0.021 &     0.027 &     0.025 &     0.031 &     0.048 &      0.101 \\
3DMolMS     &   0.039 &     0.041 &     0.057 &     0.077 &     0.070 &     0.125 &      \textbf{0.199 }\\
FixedVocab  &   0.143 &     0.184 &     \textbf{0.168 }&   0.172 &     \textbf{0.196} &     \textbf{0.220} &      0.161 \\
NEIMS (FFN) &   0.110 &     0.122 &     0.092 &     0.078 &     0.111 &     0.083 &      0.064 \\
NEIMS (GNN) &   0.155 &     0.192 &     0.164 &     \textbf{0.182} &     0.147 &     0.214 &      0.161 \\
\midrule\midrule
\ourModel       &   \textbf{0.191} &     \textbf{0.211} &     0.165 &     0.163 &     0.168 &     0.161 &      0.169 \\
\bottomrule
\end{tabular}}
\end{table}

\subsection{Dataset preparation}
\label{sec:si_datasets}

 \nistData \cite{noauthor_tandem_nodate} is prepared by extracting all positive-mode experimental spectra collected in higher-energy collision-induced dissociation (HCD) mode (i.e., collected on Orbitrap mass spectrometers). Spectra are filtered, so that we keep only those for which the associated molecule (M) has (i) a mass under 1,500 Da, (ii) contains only elements from a predefined set (i.e,. ``C'', ``N'', ``P'', ``O'', ``S'', ``Si'', ``I'', ``H'', ``Cl'', ``F'', ``Br'', ``B'', ``Se'', ``Fe'', ``Co'', ``As'', ``Na'', ``K''), and (iii) is charged with common adduct types (i.e., ``[M+H]+", ``[M+Na]+", ``[M+K]+", ``[M-H2O+]+", "[M+NH3+H]+", and "[M-2H2O+H]+"). Because non-standard empirical spectra databases~\cite{wang_sharing_2016} often do not include the measured collision energies, we pool all collision energies for each compound-adduct pairing to create a single spectrum. We refer the reader to  Young et al. ~\cite{young_massformer_2021} for detailed instructions for purchasing and extracting the \nistData dataset.

 All spectrum intensities are square-root transformed to provide higher weighting to lower intensity peaks, normalized to a maximum intensity of 1 (i.e., through dividing by the maximum intensity), filtered to exclude any noise peaks with normalized intensity under $0.003$, and subsetted to only the top 50 highest intensity peaks. All peaks are mass-shifted by the weight of the parent adduct (i.e,. if the spectrum is ``[M+H]+'', the weight of a proton is subtracted from each child peak).

\subsubsection{Product formulae assignments} 
\label{sec:subformula}
Because the
precursor ion and adduct species are known for the training dataset, we subtract
the precursor adduct mass from every peak in the training set, and attempt to
annotate each peak with a plausible product formula (i.e., a subset of the
true precursor formula). 

We opt to constrain the training product formulae to be subsets of
contiguous heavy atoms of the parent molecule as derived with the \MAGMA algorithm
\cite{ridder_automatic_2014}. 

We note two important limitations of these heuristics. First, by using molecular
substructures to annotate product formulae, our model is less prone to correctly
identifying complex rearrangements. Second, it is also possible for adduct
switching to occur. Namely, if the precursor ion has a sodium adduct
(``[M+Na]+''), some of the product formulae may actually switch and acquire a
hydrogen adduct instead. We assume no adduct switching in our formulation,
instead focusing on the novelty of the prefix tree decoding approach, as these
represent data labeling challenges, rather than modeling challenges. 

In addition, any predictive models of product formulae distributions will more closely predict spectra that would be produced on instrumentation similar to the training sets utilized  ~\cite{demarque2016fragmentation, chen2001rearrangement}. Given this, we encourage users of such models to treat these predictions as putative, rather than experimentally valid. 

\subsubsection{Dataset statistics}
\begin{figure}%
  \centering
  \includegraphics[width=\textwidth]{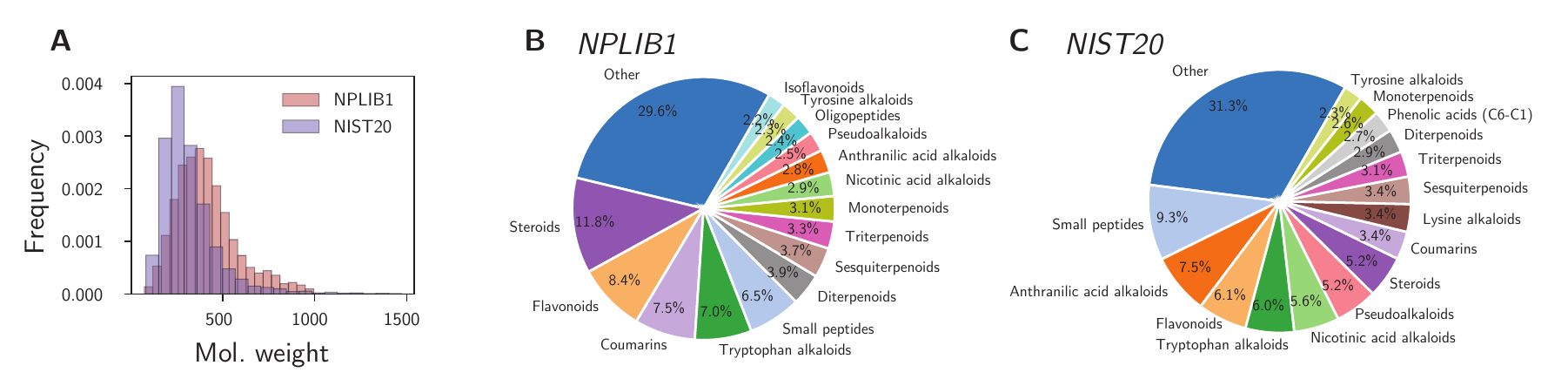}
  \caption{Spectra dataset molecule characterizations. \textbf{A.} Distribution of the molecular weight of compounds across \gnpsData and \nistData. \textbf{B-C.} Chemical classes contained in \gnpsData (B) and \nistData (C) with the top 15 classes shown and all others grouped in `Other'. Chemical classes are computed using NPClassifier \cite{kim_npclassifier_2021}.}
  \label{fig:dataset_stats}

\end{figure}

To probe the composition of our two primary datasets, we investigate both the molecular weight and chemical classes contained in the \gnpsData and \nistData datasets. We find that the average molecular weight is higher for \gnpsData (Figure \ref{fig:dataset_stats}A), consistent with the increased complexity of natural product molecules. We additionally compute chemical classes of the compounds using NPClassifier~\cite{kim_npclassifier_2021} to identify the types of compounds present in both datasets (Figure \ref{fig:dataset_stats}B-C). \gnpsData is enriched for steroids, coumarins, and various complex alkaloid natural products. On the other hand, \nistData is enriched for small peptides, nicotinic acid alkaloids, and pseudoalkaloids, among others. While these descriptions are helpful to identify the dataset composition, chemical compound classification is itself a learned classification and should be interpreted cautiously.

\subsection{Baselines}

We further describe select baseline models.

\subsubsection{CFM-ID baseline}
\label{sec:cfm}
\cfmModel\cite{allen_competitive_2015} is a long standing and important approach
to fragmentation prediction. Because \cfmModel is fit using a time intensive EM
training approach on an analogous dataset, we utilize the pretrained Docker
implementation provided by the authors in line with \cite{murphy2023efficiently}. \cfmModel has two options for predicting
molecules in either positive or negative adduct mode with ``H'' adducts (i.e.,
``[M+H]+'' or ``[M-H]-''). To directly compare to our method, we predict
spectra in positive mode and remove hydrogens from all predicted peaks, as all
training peaks are shifted by removing their adducts.

\cfmModel also produces predictions at three collision energies (i.e., low ,
medium, or high fragmentation). Because we opt not to include these, we merge
these predictions and re-normalize the result to a maximum of $1$. 

\subsubsection{Autoregressive baseline}\label{sec:autoregr}
When considering the task of generating spectrum formulae candidates, we compare \ourModelOne to an autoregressive recurrent neural network baseline, which is based around a long short term memory (LSTM) module \citep{hochreiter1997long}.

The LSTM generates formulae consecutively from a single concatenated encoding of the input molecule and input full formula. At each step in the recurrent process, a one-hot encoding of the previous predicted element count is concatenated to a one-hot encoding of the element type being predicted in the current step. By embedding this information into the network, we can avoid predicting element types that do not appear in the parent molecule's molecular formula. If the parent molecular formula has 5 element types, each autoregressively predicted formula requires generating only 5 element type counts; this eliminates the need for a stop condition between each formula. Formulae are generated autoregressively, from highest to lowest intensity. When predicting the counts of the next element type, we employ the same difference and forward count prediction strategy as used in \ourModel for fair comparison.  The model is trained with a cross entropy loss and full hyperparameters are listed in Table \ref{tab:sup_hyper_csi}.

\subsubsection{NEIMS baseline}
\label{sec:neims}
NEIMS \cite{wei_rapid_2019} is a highly efficient binned spectrum prediction
approach, originally developed for gas chromatography-mass spectra (GC-MS).
To enable a fair comparison, we optimize its hyperparameters on our dataset and add higher
resolution bins.
Furthermore, we also train a graph neural network-based version ``NEIMS (GNN)'' (in addition to the network that more closely matches \citet{wei_rapid_2019}'s original model and operates on the molecular fingerprint, ``NEIMS (FFN)''). The adduct type is either concatenated to all atom features (for NEIMS (GNN)) or to the fingerprint vector (for NEIMS (FFN)).

\subsubsection{3DMolMS baseline}
\label{sec:3dmolms}
3DMolMS \cite{hong20233dmolms} is a binned spectrum prediction
approach developed simultaneously to this work. Unlike the other binned NEIMS approach, 3DMolMS utilizes a point-based deep neural network model operating on the point cloud of an input molecule. To project a 2D molecule or SMILES string into 3D space, a single 3D conformer is first generated using RDKit \cite{landrum2016rdkit}. After several 3D convolutions, the atom-wise representations are pooled, covariates corresponding to the settings of the machine and experiment are concatenated, and the result is projected into a fixed length binned spectrum. 

To enable a fair comparison, we copy the 3DMolMS architecture into our modeling framework with minor tweaks to the network. Rather than use variable sized hidden layers, we fix hidden layer sizes to a single value across convolutions. In addition, we only use the covariate of the adduct type for consistency with our model, excluding collision energy and instrument type.  We hyperparameter optimize the model independently. We find that the performance of this model is substantially lower than the NEIMS baseline, likely due to the additional use of the ``difference'' prediction module in the NEIMS approach that allows the network to predict intensities at both fragments and neutral losses.

\subsubsection{FixedVocab baseline}
\label{sec:fixedvocab}
Concurrently to our work, \citeauthor{murphy2023efficiently} introduce an alternative formula prediction strategy for mass spectrum prediction, a model they term GRAFF-MS. Unlike \ourModel, GRAFF-MS utilizes a fixed vocabulary of molecular formulae and molecular formula differences, predicting intensities at each such value without learning to encode the formulae. These formulae and formula differences are selected in a greedy fashion based upon their frequency in the training set. 

Because no code was released for this approach at the time this work was conducted, we reimplement a variant of their approach that emphasizes the use of a fixed vocabulary of formulae and differences. For methodological consistency, we utilize equivalent formula annotations as used by \ourModel (i.e., one annotation per peak), do not model collision energies or instrument types, and utilize the same graph encoder as \ourModel for encoding each molecule. We treat the number of fragment and difference formulae as a tunable hyperparameter (which is optimized along with the rest of the hyperparameters -- see \S\ref{sec:hyps}). We mask all invalid formulae and differences and utilize a cosine similarity loss with the original spectrum to train the model. To convert predicted formula and difference intensities into a binned spectrum, each formula-intensity pairing is projected into its respective binned position using a scatter max calculation. We note that because alternative adducts and isotopes are not labeled in our preprocessing step, we do not predict isotopic or adduct variants for each fragment.

Given the differences between codebases, it is possible that the performance of our reimplementation does not exactly match the original implementation, and we instead refer to it as ``FixedVocab'' rather than GRAFF-MS in table presentations. An earlier version of our work understated the FixedVocab model's performance due to an implementation decision along these lines (specifically, not including the ``0'' neutral loss as a predicted vocabulary entry). This has since been rectified, increasing the accuracy of the FixedVocab model.

\subsection{Retrieval subsets vs. PubChem}
\label{sec:pubchem_justif}

We restrict our retrieval experiments in \S\ref{sec:retrieval} to only the top 49 decoys per test case for two reasons. First, from a practical perspective, running the forward model on every isomer match in PubChem (approx. thousands each, >1,000,000 for only 500 test cases) makes benchmarking across all considered models substantially more challenging both for this work and also future work. Second, we also believe that this top 50 challenge represents a more realistic setting. In practice, retrieval compound libraries will often be carefully crafted and designed to contain molecules similar to the unknown molecule rather than all possible isomers (using either prior knowledge or “backward” models such as CSI:FingerID \cite{duhrkop_searching_2015} and MIST \cite{Goldman2023annotating}). We conduct a side-by-side analysis on a small 500 molecule subset of the test set comparing the setting described above (with 49 decoys) to a setting with no limit on the number of decoys. The results are shown in Table \ref{tab:si_retrieval_acc_all}, showing that \ourModel still performs well in this setting.

\subsection{Model details}
\label{sec:model_details}

Here, we describe details of our model's training setup, architecture, and hyperparameters that were omitted from the main text.
Definitive details can also be found in the code \codeUrl.

\subsubsection{Training}

We train each of our models on a single RTX A5000 NVIDIA GPU (CUDA Version 11.6), making use of the 
Torch Lightning \citep{Falcon_PyTorch_Lightning_2019} library to manage the training.
\ourModelOne and \ourModelTwo take on the order of 1.5 and 2.5 hours of wall time to train respectively.

\subsubsection{Molecule encoding}
\label{sec:mol_enc}

Within both \ourModelOne and \ourModelTwo, a key component is an encoding of the molecular graph
using a message passing graph neural network, $\textsf{gnn}(\mol)$. Such graph neural network models
are now well described \citep{Hamilton2020-op,Bronstein2021-dx,Battaglia2018-im}, so we will skip a detailed explanation of them here. In our experiments, we use gated graph sequence neural networks \citep{Li2015-fc}. We made use of the implementation of this network in the DGL library \citep{Wang2019-ga} and use as atom features those shown in Table~\ref{tab:atomFeats} (which are computed using RDKit \citep{landrum2016rdkit} or DGL \citep{Wang2019-ga}).

\begin{table}[ht]
  \centering
  \caption{Graph neural network (GNN) atom features.}
  \begin{tabular}{ll}
    \toprule
    Name         & Description                            \\
    \midrule
    Element type & one-hot encoding of the element type                    \\
    Degree & one-hot encoding of number of bonds atom is associated with                    \\
    Hybridization type & \makecell{one-hot encoding of the hybridization (SP, SP2, SP3, SP3D, SP3D2)     }            \\
    Charge & one-hot encoding of atom's formal charge (from -2 to 3)  \\
    Ring-system & binary flag indicating whether atom is part of a ring \\
    Atom mass & atom's mass as a \texttt{float} \\
    Chiral tag & atom's chiral tag as one-hot encoding \\
    Adduct type  & one-hot encoding of the adduct ion                    \\
    Random walk embed steps & positional encodings of the nodes computed using DGL \\
    \bottomrule
\end{tabular}
\label{tab:atomFeats}
\end{table}

\subsubsection{Molecular formulae representations}
\label{sec:form_reps}

When forming representations of formulae (including formulae prefixes) we use a count-based encoder, $\textsf{counts}(\bm{f})$.
This encoder takes in as input the counts of all individual elements in the formula (which also can be ``undefined'' for counts of elements not yet specified -- indicated as `$\ast$' in Figure~\ref{fig:scarfModelOne}B) and returns a vector representation in $\mathbb{R}^d$.
The encoder is based upon the Fourier feature mapping proposed by \citet{Tancik2020-yy}, but using only $\sin$ basis functions (to reduce the number of parameters required by our networks).
\citet{Tancik2020-yy} has shown that such features perform better than encoding integers directly; furthermore, compared to learned representations, using Fourier features enables us (at least in principle) to deal with counts at test time that have not been seen during training.

To be precise, each possible count, $v \in \mathbb{N}_0$, is encoded by our counts-based encoder into the vector:
\begin{equation*}
\textsf{abs} \left( \left[ \sin\left( \frac{2\pi v}{T_1} \right), \sin\left( \frac{2\pi v}{T_2} \right), \sin\left( \frac{2\pi v}{T_3} \right), \ldots \right] \right),
\end{equation*}
where the periods ($T_1$, $T_2$, etc.) are set at increasing powers of two that enable us to discriminate all possible element counts given in the input, and $\textsf{abs}(\cdot)$ is the absolute value function such that we get positive embeddings.
For the ``undefined'' count we learn a separate encoding of the same dimensionality.

\subsubsection{Further details of \ourModelOne}
\label{sec:modelOne}

Pseudo-code for the \ourModelOne model is shown in Algorithm~\ref{alg:modelOne}.
Note that the second for loop (on the line marked $\ddagger$) does not depend on previous iterations of the loop, so that in practice we perform this computation in parallel.
At training time we use teacher forcing (\S \ref{sec:train}), meaning the first for loop (marked $\dagger$) is only run sequentially at inference time.

The function $\textsf{scarf-thread-net}(\cdot)$ represents the network shown in Figure~\ref{fig:scarfModelOne}B and generates the set of subsequent valid element counts given a prefix (i.e., the child nodes of a given prefix node).
As discussed in the main text, we treat this as a multi-label binary classification task and predict the binary label for each possible count using forward and difference MLPs (Eq.~\ref{eqn:fordiff}).
We fix a maximum possible element count (i.e., the number of possible classes in this classification problem), $N = 160$.
We do not allow product formulae to have more of a given element than is present in the precursor formula, $\preForm$, and we achieve this by setting the probability of these classes to zero.

\begin{algorithm}[ht]
\caption{Pseudo-code for \ourModelOne, which generates prefix trees from a root node autoregressively, one level at a time.}\label{alg:modelOne}
\KwData{Input molecule, $\mol$, with corresponding input formula, $\preForm$.}
\KwResult{Set of product formulae, $ \rho_e = \setOfProdForm$.}
\nl $\bm{h}_\mol \gets \textsf{gnn}(\mol)$  \tcp*[r]{Form embedding of precursor molecule.}
\nl $\rho_0 \gets \{ \bm{\ast} \}$  \tcp*[r]{Store the set of initial prefixes which is just the undefined formula, $\bm{\ast}$.}
\nlother{$\dagger$} \For( \tcp*[f]{Loop over all possible elements.}){$j \in [1, \ldots, e ]$}{
  \nl $\rho_j \gets \{ \, \}$ \tcp*[r]{Create the set of prefixes the next time around.}
  \nl $ \bm{h}_j \gets \textsf{one-hot}(j)$ \tcp*[r]{Encoding of which element we are predicting the count of.}
 \nlother{$\ddagger$} \ParFor(\tcp*[f]{Loop over all current prefixes.}){$\prodForm{'}_{<j} \in \rho_{j-1} $}{
    \nl $ \bm{c}' = [\bm{h}_\mol, \textsf{counts}(\prodForm{'}_{<j}), \textsf{counts}(\preForm - \prodForm{'}_{<j}), \bm{h}_j ]$ \tcp*[r]{Create context vector, Eq. \ref{eqn:contxt}}
  \nl  $ \{ \prodFormIdxd{i'}{j} \}^{n'}_{i'=1}  \gets \textsf{scarf-thread-net}(\bm{c}', \preForm) $ \tcp*[r]{Predict the set of valid next element counts under this prefix.}
  \nl  $ \rho_j \gets \rho_j \cup \textsf{create-new-prefixes}(\prodForm{'}_{<j}, \{ \prodFormIdxd{i'}{j} \}^{n'}_{i'=1} )$  \tcp*[r]{Create new prefixes for the next element.}
    
    }
  }
 \Return $\rho_e$
\end{algorithm}

\subsubsection{Further details of \ourModelTwo}
\label{sec:modelTwo}

As discussed in the main text, \ourModelTwo is based off \citet{lee2019set}'s Set Transformer.
After forming the input encoding using the molecule and count-based encoder (\S \ref{sec:mol_enc} \& \S \ref{sec:form_reps}), we further refine this embedding using an MLP (multi-layer perceptron) network.
The output of this is passed into a series of $l_3$ Transformer \citep{Vaswani2017-se} layers (\S \ref{sec:hyps} defines the exact number used in the experiments) with 8 attention heads each.

We use a cosine distance loss to train the parameters of \ourModelTwo.
This loss is also used for the FFN and GNN baselines (Table~\ref{tab:spec_acc}).
To ensure consistency with the baselines, we first project the output of
\ourModelTwo into a binned histogram representation (\S \ref{sec:hyps} defines the number of bins used); for each bin we take the max intensity across all applicable formulae.
Given a predicted binned spectra, $\predBinnedSpec$, and the ground-truth binned spectra, $\binnedSpec$, the cosine distance is defined as the negative of the cosine similarity (computed using PyTorch's \texttt{torch.cosine\_similarity} function \citep{paszke2019pytorch}):
\begin{equation}
\textsf{cos-sim}\left( \predBinnedSpec, \binnedSpec \right) = \frac{ \predBinnedSpec \cdot \binnedSpec}{ \textsf{max} \left( \| \predBinnedSpec \|_2 \| \binnedSpec \|_2, \; \epsilon \right)},
\end{equation}
where $\epsilon = \num{1e-8}$ is used to ensure numerical stability.

\subsubsection{Hyperparameters}
\label{sec:hyps}

To enable fair comparison across models, hyperparameters were tuned for
\ourModel, the FFN binned prediction baseline, and the GNN binned prediction
baseline.  Parameters were tuned using RayTune \cite{liaw2018tune} with Optuna \cite{akiba2019optuna} and an
ASHAScheduler. Each model was allotted 50 different hyperoptimization trials for
fitting. Models were hyperparameter optimized on a smaller $10,000$ spectra subset
of \nistData.  Parameters are detailed in Table \ref{tab:sup_hyper_csi}.

\subsection{Limitations and future work}
\label{sec:future_work}

We outline several potential directions for future work to address limitations of this work. 

\begin{enumerate}
    \item \emph{Improving the gold standard training annotation pre-processing.} Because \ourModel is flexible in that it can match distributions of formula assignments, a key step to improving and building upon this approach is to develop more robust assignments of formula to training spectra. This includes adding complexity and removing potential assumptions, such as allowing annotations to account for rearrangement, elimination, or charge transfer. A second goal is to identify potentially low quality training spectra, such as ones that emerge from mixtures, and remove these from the inputs. Another potential way to handle such cases would be to model each spectrum peak as an \emph{ensemble} of potential equivalent-mass formulae, which would be particularly helpful in relating \ourModel to inverse models such as MIST \cite{goldman_generating_2023} in which the structure of the molecule and formula identity of each peak cannot be known \emph{a priori}.
    \item \emph{Incorporating other model covariates.} Incorporating collision energy features explicitly into the model, as well as negative-ion mode inputs, will increase its usability. This could be enabled by aggregating public data containing these annotations.
    \item \emph{Featurizing molecule inputs using different or more powerful molecular encoders.} Recent and simultaneous work to this used a pretrained graph encoder as part of a binned spectrum prediction approach, MassFormer~\cite{young_massformer_2021}. It is possible to include more powerful molecule or formula encoders into \ourModel.
    \item \emph{Consideration of interpretability by subgraph attribution and combination with ICEBERG.} %
    Following this initial work, we developed a second model, ICEBERG \cite{goldman_generating_2023}, that uses a similar two step modeling approach, but instead encodes fragments, not formula. This increases accuracy and robustness, especially for retrieval, but substantially slows the model. In comparison to ICEBERG, \ourModel still has several benefits including speed, the lack of required substructure labeling, and ability to capture potential skeletal rearrangements of molecules (i.e., discontinuities in structure that may not be possible to model by only breaking bonds). An open question and exciting opportunity in the future is to combine these two levels of abstraction and make formula-level predictions with graph-level attribution or featurization. 

    \item \emph{Retrieval-specific loss functions to enhance retrieval performance.} A significant finding of this work was the noise associated with the retrieval task and lack of correlation with spectrum prediction performance as measured by cosine similarity. Future work may consider how to more directly define loss functions that reflect the task of retrieval.
\end{enumerate}

\begingroup
\small
\setlength{\tabcolsep}{4pt} %
\begin{centering}
\begin{longtable}{@{}llll@{}}
\caption{Model and baseline hyperparameters.}
\label{tab:sup_hyper_csi}\\
\toprule
\textbf{Model} & \textbf{Parameter}                               & \textbf{Grid}               & \textbf{Value}
\endfirsthead
\multicolumn{4}{c}%
{{\bfseries \tablename\ \thetable{} -- continued from previous page}} \\
\midrule \textbf{Model} & \textbf{Parameter}                               & \textbf{Grid}               & \textbf{Value}\\\midrule
\endhead
\midrule
Autoregressive    & learning rate                                    & $[1e-4,1e-3]$               & $0.0009$      \\
               & learning rate scheduler                                        & \--               & StepDecay (5,000)          \\
               & learning rate decay                               & $[0.7, 1.0]$              & 0.85            \\
               & dropout                                          & $\{0.0, 0.1,0.2,0.3\}$           & $0.2$          \\
               & hidden size, $d$                                 & $\{128,256,512\}$        & $512$          \\
               & gnn layers                                      & $[1,6]$                 & $1$            \\
               & rnn layers                                      & $[1,3]$                 & $3$            \\
               & batch size                                      & $\{8,16,32,64\}$                 & $64$            \\
               & weight decay                                      & $\{0, 1e-6, 1e-7\}$                 & $1e-6$            \\
               & use differences  (Eq.\ref{eqn:fordiff})                                     & \{True, False\}                 & True            \\
               & conv type                                       & \--                         & GatedGraphConv           \\

               & random walk embed steps  (Table~\ref{tab:atomFeats})                                     & [0,20]                         & 20           \\
               & graph pooling                                       & \{mean, attention\}                         & mean           \\
\midrule
NEIMS (FFN)    & learning rate                                    & $[1e-4,1e-3]$               & $0.00087$      \\
               & learning rate scheduler                                        & \--               & StepDecay (5,000)          \\
               & learning rate decay                               & $[0.7, 1.0]$              & 0.722            \\
               & dropout                                          & $\{0.0, 0.1,0.2,0.3\}$           & $0.0$          \\
               & hidden size, $d$                                 & $\{64,128,256,512\}$        & $512$          \\
               & layers, $l$                                      & $\{1,2,3\}$                 & $2$            \\
               & batch size                                      & $\{16, 32, 64, 128\}$                 & $128$            \\
               & weight decay                                      & $\{0, 1e-6, 1e-7\}$                 & $0$            \\
               & use differences  (Eq.\ref{eqn:fordiff})                                     & \{True, False\}                 & True            \\
               & num bins  (\S \ref{sec:modelTwo})                                        & \--                         & $15,000$           \\

\midrule
NEIMS (GNN)            & learning rate                                    & $[1e-4,1e-3]$               & $0.00052$      \\
               & learning rate scheduler                                        & \--               & StepDecay (5,000)          \\
               & learning rate decay                               & $[0.7, 1.0]$              & 0.767            \\
               & dropout                                          & $\{0.0, 0.1,0.2,0.3\}$           & $0.0$          \\
               & hidden size, $d$                                 & $\{64,128,256, 512\}$        & $512$          \\
               & layers, $l$                                      & $[1,6]$                 & $4$            \\
               & batch size                                      & $\{16, 32, 64\}$                 & $64$            \\
               & weight decay                                      & $\{0, 1e-6, 1e-7\}$                 & $1e-7$            \\
               & use differences (Eq.\ref{eqn:fordiff})                                     & \{True, False\}                 & True            \\
               & num bins (\S \ref{sec:modelTwo})                                         & \--                         & $15,000$           \\
               & conv type                                       & \--                         & GatedGraphConv           \\

               & random walk embed steps  (Table~\ref{tab:atomFeats})                                     & [0,20]                         & 19           \\
               & graph pooling                                       & \{mean, attention\}                         & mean           \\
\midrule
3DMolMS            & learning rate                                    & $[1e-4,1e-3]$               & $0.00074$      \\
               & learning rate scheduler                                        & \--               & StepDecay (5,000)          \\
               & learning rate decay                               & $[0.7, 1.0]$              & 0.86            \\
               & dropout                                          & $\{0.0, 0.1,0.2,0.3\}$           & $0.3$          \\
               & hidden size, $d$                                 & $\{64,128,256, 512\}$        & $256$          \\
               & layers, $l$                                      & $[1,6]$                 & $2$            \\
               & top layers                                      & $[1,3]$                 & $2$            \\
               & neighbors, $k$                                      & $[3,6]$                 & $5$            \\
               & batch size                                      & $\{16, 32, 64\}$                 & $16$            \\
               & weight decay                                      & $\{0, 1e-6, 1e-7\}$                 & $1e-6$            \\
               & num bins (\S \ref{sec:modelTwo})                                         & \--                         & $15,000$           \\
\midrule
FixedVocab            & learning rate                                    & $[1e-4,1e-3]$               & $0.00018$      \\
               & learning rate scheduler                                        & \--               & StepDecay (5,000)          \\
               & learning rate decay                               & $[0.7, 1.0]$              & 0.92            \\
               & dropout                                          & $\{0.0, 0.1,0.2,0.3\}$           & $0.3$          \\
               & hidden size, $d$                                 & $\{64,128,256, 512\}$        & $512$          \\
               & layers, $l$                                      & $[1,6]$                 & $6$            \\
               & batch size                                      & $\{16, 32, 64\}$                 & $64$            \\
               & weight decay                                      & $\{0, 1e-6, 1e-7\}$                 & $1e-6$            \\
               & num bins (\S \ref{sec:modelTwo})                                         & \--                         & $15,000$           \\
                & conv type                                       & \--                         & GatedGraphConv           \\
                
               & random walk embed steps  (Table~\ref{tab:atomFeats})                                     & [0,20]                         & 11           \\
               & graph pooling                                       & \{mean, attention\}                         & mean   \\
              & formula library size                                       & \{1000, 5000, 10000, 25000, 50000\}                         & 5000   \\
\midrule
\midrule
\ourModelOne   & learning rate                                    & $[1e-4,1e-3]$               & $0.000577$      \\
               & learning rate scheduler                                        & \--               & StepDecay (5,000)          \\
               & learning rate decay                               & $[0.7, 1.0]$              & 0.894            \\
               & dropout                                          & $\{0.0, 0.1,0.2,0.3\}$           & $0.3$          \\
               & hidden size, $d$                                 & $\{128, 256, 512\}$        & $512$          \\
               & mlp layers, $l_1$                                      & $[1,3]$                 & $2$            \\
               & gnn layers, $l_2$    (\S \ref{sec:mol_enc})                                    & $[1,6]$                 & $4$            \\
               & batch size                                      & $\{8, 16, 32, 64\}$                 & $16$            \\
               & weight decay                                      & $\{0, 1e-6, 1e-7\}$                 & $1e-6$            \\
               & use differences  (Eq.\ref{eqn:fordiff})                                      & \{True, False\}                 & True            \\
               & conv type                                       & \--                         & GatedGraphConv           \\
               & random walk embed steps  (Table~\ref{tab:atomFeats})                                       & [0,20]                         & 20           \\
               & graph pooling                                       & \{mean, attention\}                         & mean           \\

\midrule
\ourModelTwo   & learning rate                                    & $[1e-4,1e-3]$               & $0.00031$      \\
               & learning rate scheduler                                        & \--               & StepDecay (5,000)          \\
               & learning rate decay                               & $[0.7, 1.0]$              & 0.962            \\
               & dropout                                          & $\{0.0, 0.1,0.2,0.3\}$           & $0.2$          \\
               & hidden size, $d$                                 & $\{128, 256, 512\}$        & $512$          \\
               & mlp layers, $l_1$   (\S \ref{sec:modelTwo})                                    & $[1,3]$                 & $2$            \\
               & gnn layers, $l_2$   (\S \ref{sec:mol_enc})                                   & $[1,6]$                 & $3$            \\
               & transformer layers, $l_3$ (\S \ref{sec:modelTwo})                                     & $[0,3]$                 & $2$            \\
               & batch size                                      & $\{4, 8, 16, 32, 64\}$                 & $32$            \\
               & weight decay                                      & $\{0, 1e-6, 1e-7\}$                 & $0$            \\
               & num bins   (\S \ref{sec:modelTwo})                                    & \--                         & $15,000$           \\
               & conv type                                       & \--                         & GatedGraphConv           \\
               & random walk embed steps   (Table~\ref{tab:atomFeats})                                      & [0,20]                         & 7           \\
               & graph pooling                                       & \{mean, attention\}                         & attention           \\

\midrule

\end{longtable}
\end{centering}
\endgroup  

\FloatBarrier

\end{document}